\documentclass{eptcs}
\usepackage{isabelle}
\usepackage{isabellesym}
\usepackage{color}
\usepackage{amsthm}
\newtheorem{theorem}{Theorem}
\newtheorem{lemma}[theorem]{Lemma}
\usepackage{amsmath}
\usepackage{amssymb} 
\usepackage{ot1patch}
\usepackage{times}
\usepackage{proof}
\usepackage{mathpartir}

\isabellestyle{it}
\renewcommand{\isastyleminor}{\it}%
\renewcommand{\isastyle}{\normalsize\it}%

\def\dn{\,\stackrel{\mbox{\scriptsize def}}{=}\,}
\renewcommand{\isasymequiv}{$\dn$}
\renewcommand{\isasymemptyset}{$\varnothing$}

\def\titlerunning{Nominal Unification Revisited}
\def\authorrunning{C.~Urban}
\definecolor{mygrey}{rgb}{.80,.80,.80}

\begin{document}

\title{Nominal Unification Revisited}
\author{Christian Urban 
\institute{TU Munich, Germany}
\email{urbanc@in.tum.de}}
\maketitle

\begin{abstract} 
  Nominal unification calculates substitutions that make terms involving
  binders equal modulo alpha-equivalence. Although nominal unification can be seen as
  equivalent to Miller's higher-order pattern unification, it has 
  properties, such as the use of first-order terms with names (as opposed to
  alpha-equivalence classes) and that no new names need to be generated during
  unification, which set it clearly apart from higher-order pattern unification. The
  purpose of this paper is to simplify a clunky proof from the original
  paper on nominal unification and to give an overview over some results 
about nominal unification.
\end{abstract}

\begin{isabellebody}%
\def\isabellecontext{Paper}%
\isadelimtheory
\endisadelimtheory
\isatagtheory
\endisatagtheory
{\isafoldtheory}%
\isadelimtheory
\endisadelimtheory
\isamarkupsection{Introduction%
}
\isamarkuptrue%
\begin{isamarkuptext}%
The well-known first-order unification algorithm by Robinson \cite{Robinson65} 
  calculates substitutions for variables that make terms syntactically equal. For 
  example the terms 

  \begin{center}
  \isa{f\ {\isaliteral{5C3C6C616E676C653E}{\isasymlangle}}X{\isaliteral{2C}{\isacharcomma}}\ X{\isaliteral{5C3C72616E676C653E}{\isasymrangle}}\ {\isaliteral{3D}{\isacharequal}}\isaliteral{5C3C5E7375703E}{}\isactrlsup {\isaliteral{3F}{\isacharquery}}\ f\ {\isaliteral{5C3C6C616E676C653E}{\isasymlangle}}Z{\isaliteral{2C}{\isacharcomma}}\ g\ {\isaliteral{5C3C6C616E676C653E}{\isasymlangle}}Y{\isaliteral{5C3C72616E676C653E}{\isasymrangle}}{\isaliteral{5C3C72616E676C653E}{\isasymrangle}}}
  \end{center}

  \noindent
  can be made syntactically equal with the substitution \isa{{\isaliteral{5B}{\isacharbrackleft}}X\ {\isaliteral{3A}{\isacharcolon}}{\isaliteral{3D}{\isacharequal}}g\ {\isaliteral{5C3C6C616E676C653E}{\isasymlangle}}Y{\isaliteral{5C3C72616E676C653E}{\isasymrangle}}{\isaliteral{2C}{\isacharcomma}}\ Z\ {\isaliteral{3A}{\isacharcolon}}{\isaliteral{3D}{\isacharequal}}\ g\ {\isaliteral{5C3C6C616E676C653E}{\isasymlangle}}Y{\isaliteral{5C3C72616E676C653E}{\isasymrangle}}{\isaliteral{5D}{\isacharbrackright}}}.  In first-order unification we can regard variables as
  ``holes'' for which the unification algorithm calculates terms with which
  the holes need be ``filled'' by substitution. The filling
  operation is a simple replacement of terms for variables.  However, when binders come
  into play, this simple picture becomes more complicated: We are no 
  longer interested in syntactic equality since terms like

  \begin{equation}\label{ex1}
  \isa{a{\isaliteral{2E}{\isachardot}}{\isaliteral{5C3C6C616E676C653E}{\isasymlangle}}a{\isaliteral{2C}{\isacharcomma}}\ c{\isaliteral{5C3C72616E676C653E}{\isasymrangle}}\ {\isaliteral{5C3C617070726F783E}{\isasymapprox}}\isaliteral{5C3C5E7375703E}{}\isactrlsup {\isaliteral{3F}{\isacharquery}}\ b{\isaliteral{2E}{\isachardot}}{\isaliteral{5C3C6C616E676C653E}{\isasymlangle}}X{\isaliteral{2C}{\isacharcomma}}\ c{\isaliteral{5C3C72616E676C653E}{\isasymrangle}}}
  \end{equation}

  \noindent
  should unify, despite the fact that the binders \isa{a} and \isa{b}
  disagree. (Following \cite{UrbanPittsGabbay04} we write \isa{a{\isaliteral{2E}{\isachardot}}t} for
  the term where the name \isa{a} is bound in \isa{t}, and \isa{{\isaliteral{5C3C6C616E676C653E}{\isasymlangle}}t\isaliteral{5C3C5E697375623E}{}\isactrlisub {\isadigit{1}}{\isaliteral{2C}{\isacharcomma}}\ t\isaliteral{5C3C5E697375623E}{}\isactrlisub {\isadigit{2}}{\isaliteral{5C3C72616E676C653E}{\isasymrangle}}} 
  for a pair of terms.) If we replace \isa{X} 
  with term \isa{b} in \eqref{ex1} we obtain the instance

  \begin{equation}\label{ex2}
  \isa{a{\isaliteral{2E}{\isachardot}}{\isaliteral{5C3C6C616E676C653E}{\isasymlangle}}a{\isaliteral{2C}{\isacharcomma}}\ c{\isaliteral{5C3C72616E676C653E}{\isasymrangle}}\ {\isaliteral{5C3C617070726F783E}{\isasymapprox}}\ b{\isaliteral{2E}{\isachardot}}{\isaliteral{5C3C6C616E676C653E}{\isasymlangle}}b{\isaliteral{2C}{\isacharcomma}}\ c{\isaliteral{5C3C72616E676C653E}{\isasymrangle}}}
  \end{equation}

  \noindent
  which are indeed two alpha-equivalent terms. Therefore in a setting with binders,
  unification has to be modulo alpha-equivalence.

  What is interesting about nominal unification
  is the fact that it maintains the view from first-order unification of a variable being a 
  ``hole'' into which a term can be filled. As can be seen, by going from \eqref{ex1}
  to \eqref{ex2} we are replacing \isa{X} with the term \isa{b} without bothering
  that this \isa{b} will become bound by the binder. This means 
  the operation of substitution in nominal
  unification is possibly capturing. A result is that many complications stemming from
  the fact that binders need to be renamed when a capture-avoiding substitution is 
  pushed under a binder do not apply
  to nominal unification. Its definition of substitution states that in case of binders

  \begin{center}
  \isa{{\isaliteral{5C3C7369676D613E}{\isasymsigma}}{\isaliteral{28}{\isacharparenleft}}a{\isaliteral{2E}{\isachardot}}t{\isaliteral{29}{\isacharparenright}}\ {\isaliteral{3D}{\isacharequal}}\ a{\isaliteral{2E}{\isachardot}}{\isaliteral{5C3C7369676D613E}{\isasymsigma}}{\isaliteral{28}{\isacharparenleft}}t{\isaliteral{29}{\isacharparenright}}}
  \end{center}

  \noindent
  holds without any side-condition about \isa{a} and \isa{{\isaliteral{5C3C7369676D613E}{\isasymsigma}}}. In order to 
  obtain a unification algorithm 
  that, roughly speaking, preserves alpha-equivalence,
  nominal unification uses the notion of freshness of a name for a term. This will be 
  written as the judgement \isa{a\ {\isaliteral{23}{\isacharhash}}\ t}. For example in
  \eqref{ex1} it is ensured that the bound name \isa{a} 
  on the left-hand side is fresh for the term on the right-hand side, that means it cannot occur free
  on the right-hand side. In general
  two abstraction terms will \emph{not} unify, if the binder form one side is free
  on the other. This condition is sufficient to ensure that unification preserves alpha-equivalence
  and allows us to regard variables as holes with a simple substitution operation to fill them.

  Whenever two abstractions with different binders need to be unified, nominal
  unification uses the operation of swapping two names to rename the bound
  names. For example when solving the problem shown in \eqref{ex1}, which has
  two binders whose names disagree, then it will attempt to unify the bodies
  \isa{{\isaliteral{5C3C6C616E676C653E}{\isasymlangle}}a{\isaliteral{2C}{\isacharcomma}}\ c{\isaliteral{5C3C72616E676C653E}{\isasymrangle}}} and \isa{{\isaliteral{5C3C6C616E676C653E}{\isasymlangle}}X{\isaliteral{2C}{\isacharcomma}}\ c{\isaliteral{5C3C72616E676C653E}{\isasymrangle}}}, but first applies the swapping \isa{{\isaliteral{28}{\isacharparenleft}}a\ b{\isaliteral{29}{\isacharparenright}}} to \isa{{\isaliteral{5C3C6C616E676C653E}{\isasymlangle}}X{\isaliteral{2C}{\isacharcomma}}\ c{\isaliteral{5C3C72616E676C653E}{\isasymrangle}}}. While it is easy to see how this swapping
  should affect the name \isa{c} (namely not at all), the interesting
  question is how this swapping should affect the variable \isa{X}? Since
  variables are holes for which nothing is known until they are substituted
  for, the answer taken in nominal unification is to suspend such swapping in
  front of variables. Several such swapping can potentially accumulate in
  front of variables. In the example above, this means applying the swapping
  \isa{{\isaliteral{28}{\isacharparenleft}}a\ b{\isaliteral{29}{\isacharparenright}}} to \isa{{\isaliteral{5C3C6C616E676C653E}{\isasymlangle}}X{\isaliteral{2C}{\isacharcomma}}\ c{\isaliteral{5C3C72616E676C653E}{\isasymrangle}}} gives the term \isa{{\isaliteral{5C3C6C616E676C653E}{\isasymlangle}}{\isaliteral{28}{\isacharparenleft}}a\ b{\isaliteral{29}{\isacharparenright}}{\isaliteral{5C3C62756C6C65743E}{\isasymbullet}}X{\isaliteral{2C}{\isacharcomma}}\ c{\isaliteral{5C3C72616E676C653E}{\isasymrangle}}}, where \isa{{\isaliteral{28}{\isacharparenleft}}a\ b{\isaliteral{29}{\isacharparenright}}} is suspended in front of \isa{X}. 
  The substitution \isa{{\isaliteral{5B}{\isacharbrackleft}}X\ {\isaliteral{3A}{\isacharcolon}}{\isaliteral{3D}{\isacharequal}}\ b{\isaliteral{5D}{\isacharbrackright}}} is then determined by unifying
  the first components of the two pairs, namely \isa{a\ {\isaliteral{5C3C617070726F783E}{\isasymapprox}}\isaliteral{5C3C5E7375703E}{}\isactrlsup {\isaliteral{3F}{\isacharquery}}\ {\isaliteral{28}{\isacharparenleft}}a\ b{\isaliteral{29}{\isacharparenright}}{\isaliteral{5C3C62756C6C65743E}{\isasymbullet}}X}. 
  We can extract the substitution by applying the swapping to the term \isa{a}, giving
  \isa{{\isaliteral{5B}{\isacharbrackleft}}X\ {\isaliteral{3A}{\isacharcolon}}{\isaliteral{3D}{\isacharequal}}\ b{\isaliteral{5D}{\isacharbrackright}}}.
  This 
  method of suspending swappings in
  front of variables is related to unification in explicit substitution
  calculi which use de Bruijn indices and which record explicitly when indices must
  be raised \cite{Dowek96}.

  Nominal unification gives a similar answer to the problem of deciding when a name is fresh
  for a term containing variables, say \isa{a\ {\isaliteral{23}{\isacharhash}}\ {\isaliteral{5C3C6C616E676C653E}{\isasymlangle}}X{\isaliteral{2C}{\isacharcomma}}\ c{\isaliteral{5C3C72616E676C653E}{\isasymrangle}}}. In this case 
  it will record explicitly that \isa{a} must be fresh for \isa{X}. (Since we assume 
  \isa{a\ {\isaliteral{5C3C6E6F7465713E}{\isasymnoteq}}\ c}, it will be that \isa{a} is fresh for \isa{c}.) 
  This amounts to
  the constraint that  nothing can be substituted for \isa{X} that contains a free occurrence
  of \isa{a}. Consequently the judgements for freshness \isa{{\isaliteral{23}{\isacharhash}}}, and also equality \isa{{\isaliteral{5C3C617070726F783E}{\isasymapprox}}}, 
  depend on an explicit freshness context recording what variables need to be fresh for. 
  We will give the inductive definitions for \isa{{\isaliteral{23}{\isacharhash}}} and \isa{{\isaliteral{5C3C617070726F783E}{\isasymapprox}}} in Section~\ref{mainsec}.
  This method of recording extra freshness constraints also allows us to regard
  the following two terms containing a hole (the variable \isa{X})

  \begin{center}
  \isa{a{\isaliteral{2E}{\isachardot}}X\ {\isaliteral{5C3C617070726F783E}{\isasymapprox}}\ b{\isaliteral{2E}{\isachardot}}X}
  \end{center}
  
  \noindent
  as alpha-equal---namely under the condition that both \isa{a} and \isa{b} must
  be fresh for the variable \isa{X}. This is defined in terms of judgements of the form

  \begin{center}
  \isa{{\isaliteral{7B}{\isacharbraceleft}}a\ {\isaliteral{23}{\isacharhash}}\ X{\isaliteral{2C}{\isacharcomma}}\ b\ {\isaliteral{23}{\isacharhash}}\ X{\isaliteral{7D}{\isacharbraceright}}\ {\isaliteral{5C3C7475726E7374696C653E}{\isasymturnstile}}\ a{\isaliteral{2E}{\isachardot}}X\ {\isaliteral{5C3C617070726F783E}{\isasymapprox}}\ b{\isaliteral{2E}{\isachardot}}X}
  \end{center}

  \noindent
  The reader can easily determine that any substitution 
  for \isa{X} that satisfies these freshness conditions will produce two
  alpha-equivalent terms. 

  Unification problems solved by nominal unification occur frequently in practice. 
  For example typing rules are typically specified as:

  \begin{center}
  \begin{tabular}{c}
  \infer[var]{\isa{{\isaliteral{5C3C47616D6D613E}{\isasymGamma}}\ {\isaliteral{5C3C7475726E7374696C653E}{\isasymturnstile}}\ x\ {\isaliteral{3A}{\isacharcolon}}\ {\isaliteral{5C3C7461753E}{\isasymtau}}}}{\isa{{\isaliteral{28}{\isacharparenleft}}x{\isaliteral{2C}{\isacharcomma}}\ {\isaliteral{5C3C7461753E}{\isasymtau}}{\isaliteral{29}{\isacharparenright}}\ {\isaliteral{5C3C696E3E}{\isasymin}}\ {\isaliteral{5C3C47616D6D613E}{\isasymGamma}}}}\hspace{7mm}
  \infer[app]{\isa{{\isaliteral{5C3C47616D6D613E}{\isasymGamma}}\ {\isaliteral{5C3C7475726E7374696C653E}{\isasymturnstile}}\ t\isaliteral{5C3C5E697375623E}{}\isactrlisub {\isadigit{1}}\ t\isaliteral{5C3C5E697375623E}{}\isactrlisub {\isadigit{2}}\ {\isaliteral{3A}{\isacharcolon}}\ {\isaliteral{5C3C7461753E}{\isasymtau}}\isaliteral{5C3C5E697375623E}{}\isactrlisub {\isadigit{2}}}}{\isa{{\isaliteral{5C3C47616D6D613E}{\isasymGamma}}\ {\isaliteral{5C3C7475726E7374696C653E}{\isasymturnstile}}\ t\isaliteral{5C3C5E697375623E}{}\isactrlisub {\isadigit{1}}\ {\isaliteral{3A}{\isacharcolon}}\ {\isaliteral{5C3C7461753E}{\isasymtau}}\isaliteral{5C3C5E697375623E}{}\isactrlisub {\isadigit{1}}\ {\isaliteral{5C3C72696768746172726F773E}{\isasymrightarrow}}\ {\isaliteral{5C3C7461753E}{\isasymtau}}\isaliteral{5C3C5E697375623E}{}\isactrlisub {\isadigit{2}}} & \isa{{\isaliteral{5C3C47616D6D613E}{\isasymGamma}}\ {\isaliteral{5C3C7475726E7374696C653E}{\isasymturnstile}}\ t\isaliteral{5C3C5E697375623E}{}\isactrlisub {\isadigit{2}}\ {\isaliteral{3A}{\isacharcolon}}\ {\isaliteral{5C3C7461753E}{\isasymtau}}\isaliteral{5C3C5E697375623E}{}\isactrlisub {\isadigit{1}}}}\hspace{7mm}
  \infer[lam]{\isa{{\isaliteral{5C3C47616D6D613E}{\isasymGamma}}\ {\isaliteral{5C3C7475726E7374696C653E}{\isasymturnstile}}\ {\isaliteral{5C3C6C616D6264613E}{\isasymlambda}}x{\isaliteral{2E}{\isachardot}}t\ {\isaliteral{3A}{\isacharcolon}}\ {\isaliteral{5C3C7461753E}{\isasymtau}}\isaliteral{5C3C5E697375623E}{}\isactrlisub {\isadigit{1}}\ {\isaliteral{5C3C72696768746172726F773E}{\isasymrightarrow}}\ {\isaliteral{5C3C7461753E}{\isasymtau}}\isaliteral{5C3C5E697375623E}{}\isactrlisub {\isadigit{2}}}}{\isa{{\isaliteral{28}{\isacharparenleft}}x{\isaliteral{2C}{\isacharcomma}}\ {\isaliteral{5C3C7461753E}{\isasymtau}}\isaliteral{5C3C5E697375623E}{}\isactrlisub {\isadigit{1}}{\isaliteral{29}{\isacharparenright}}{\isaliteral{3A}{\isacharcolon}}{\isaliteral{3A}{\isacharcolon}}{\isaliteral{5C3C47616D6D613E}{\isasymGamma}}\ {\isaliteral{5C3C7475726E7374696C653E}{\isasymturnstile}}\ t\ {\isaliteral{3A}{\isacharcolon}}\ {\isaliteral{5C3C7461753E}{\isasymtau}}\isaliteral{5C3C5E697375623E}{}\isactrlisub {\isadigit{2}}} & \isa{x\ {\isaliteral{5C3C6E6F74696E3E}{\isasymnotin}}\ dom\ {\isaliteral{5C3C47616D6D613E}{\isasymGamma}}}}
  \end{tabular}
  \end{center}

  \noindent
  Assuming we have the typing judgement \isa{{\isaliteral{5C3C656D7074797365743E}{\isasymemptyset}}\ {\isaliteral{5C3C7475726E7374696C653E}{\isasymturnstile}}\ {\isaliteral{5C3C6C616D6264613E}{\isasymlambda}}y{\isaliteral{2E}{\isachardot}}s\ {\isaliteral{3A}{\isacharcolon}}\ {\isaliteral{5C3C7369676D613E}{\isasymsigma}}}, we are 
  interested how the \isa{lam}-rule, the only one that unifies, needs to be instantiated in order to
  derive the premises under which \isa{{\isaliteral{5C3C6C616D6264613E}{\isasymlambda}}y{\isaliteral{2E}{\isachardot}}s} is typable. This leads to the
  nominal unification problem

  \begin{center}
  \isa{{\isaliteral{5C3C656D7074797365743E}{\isasymemptyset}}\ {\isaliteral{5C3C7475726E7374696C653E}{\isasymturnstile}}\ {\isaliteral{5C3C6C616D6264613E}{\isasymlambda}}y{\isaliteral{2E}{\isachardot}}s\ {\isaliteral{3A}{\isacharcolon}}\ {\isaliteral{5C3C7369676D613E}{\isasymsigma}}\ {\isaliteral{5C3C617070726F783E}{\isasymapprox}}\isaliteral{5C3C5E7375703E}{}\isactrlsup {\isaliteral{3F}{\isacharquery}}\ {\isaliteral{5C3C47616D6D613E}{\isasymGamma}}\ {\isaliteral{5C3C7475726E7374696C653E}{\isasymturnstile}}\ {\isaliteral{5C3C6C616D6264613E}{\isasymlambda}}x{\isaliteral{2E}{\isachardot}}t\ {\isaliteral{3A}{\isacharcolon}}\ {\isaliteral{5C3C7461753E}{\isasymtau}}\isaliteral{5C3C5E697375623E}{}\isactrlisub {\isadigit{1}}\ {\isaliteral{5C3C72696768746172726F773E}{\isasymrightarrow}}\ {\isaliteral{5C3C7461753E}{\isasymtau}}\isaliteral{5C3C5E697375623E}{}\isactrlisub {\isadigit{2}}}
  \end{center}

  \noindent
  which can be solved by the substitution \isa{{\isaliteral{5B}{\isacharbrackleft}}{\isaliteral{5C3C47616D6D613E}{\isasymGamma}}\ {\isaliteral{3A}{\isacharcolon}}{\isaliteral{3D}{\isacharequal}}\ {\isaliteral{5C3C656D7074797365743E}{\isasymemptyset}}{\isaliteral{2C}{\isacharcomma}}\ t\ {\isaliteral{3A}{\isacharcolon}}{\isaliteral{3D}{\isacharequal}}\ {\isaliteral{28}{\isacharparenleft}}y\ x{\isaliteral{29}{\isacharparenright}}\ {\isaliteral{5C3C62756C6C65743E}{\isasymbullet}}\ s{\isaliteral{2C}{\isacharcomma}}\ {\isaliteral{5C3C7369676D613E}{\isasymsigma}}\ {\isaliteral{3A}{\isacharcolon}}{\isaliteral{3D}{\isacharequal}}\ {\isaliteral{5C3C7461753E}{\isasymtau}}\isaliteral{5C3C5E697375623E}{}\isactrlisub {\isadigit{1}}\ {\isaliteral{5C3C72696768746172726F773E}{\isasymrightarrow}}\ {\isaliteral{5C3C7461753E}{\isasymtau}}\isaliteral{5C3C5E697375623E}{}\isactrlisub {\isadigit{2}}{\isaliteral{5D}{\isacharbrackright}}} with 
  the requirement that \isa{x} needs to be fresh for \isa{s} (in order to stay close to informal 
  practice, we deviate here from the convention of using upper-case letters for variables and 
  lower-case letters for names). 

  Most closely related to nominal unification is higher-order pattern unification by 
  Miller \cite{Miller91}. Indeed Cheney has shown that higher-order pattern unification problems
  can be solved by an encoding to nominal unification problems \cite{cheney05}. Levy and Villaret 
  have  presented an encoding for the other direction \cite{levyvillaret08}. However, there are 
  crucial differences between both methods of unifying terms with binders. One difference is that 
  nominal unification regards variables as holes for which terms can be 
  substituted in a possibly
  capturing manner. In contrast, higher-order pattern unification is based on the notion of 
  capture-avoiding substitutions. Hence, variables are not just holes, but always need to come 
  with the parameters, or names, the variable 
  may depend on. For example in order to imitate the behaviour of \eqref{ex1}, we have to
  write \isa{X\ a\ b}, explicitly indicating that the variable \isa{X} may depend 
  on \isa{a} and \isa{b}. If we replace \isa{X} with an appropriate lambda-abstraction, then 
  the dependency can by ``realised'' via a beta-reduction.
  This results in unification problems involving lambda-terms to be unified
  modulo alpha, beta and eta equivalence. In order to make this kind of unification 
  problems to be decidable, Miller introduced restrictions on the form of the lambda-terms to be unified. 
  With this restriction he obtains unification problems that are not only decidable, but 
  also possess (if solvable) most general solutions.
  
  Another difference between nominal unification and higher-order pattern unification is that the 
  former uses first-order terms, while the latter uses alpha-equivalence classes. This makes the
  implementation of higher-order pattern unification in a programming language
  like ML substantially harder than an implementation of nominal
  unification. One possibility is to implement elements of alpha-equivalence classes as 
  trees and then be very careful in the treatment
  of names, generating new ones on the fly. Another 
  possibility is to implement them with 
  de-Bruijn indices. Both possibilities, unfortunately,
  give rise to rather complicated unification algorithms. This 
  complexity is one reason that higher-order
  unification has up to now not been formalised in a theorem prover, whereas nominal
  unification has been formalised twice \cite{UrbanPittsGabbay04, KumarNorrish10}. 
  One concrete example for the higher-order
  pattern unification algorithm being more complicated than the 
  nominal unification algorithm is the following: higher-order 
  pattern unification has been part of the infrastructure of the 
  Isabelle theorem prover for many years
  \cite{Nipkow93}. The problem, unfortunately,  with this implementation is that it unifies 
  a slightly enriched term-language (which allows general beta-redexes) and it is not completely
  understood how eta and beta equality interact in this algorithm. A formalisation
  of Isabelle's version of higher-order pattern unification and its claims is 
  therefore very
  much desired, since any bug can potentially compromise
  the correctness of Isabelle.

  In a formalisation it is important to have the simplest possible argument
  for establishing a property, since this nearly always yields a simple
  formalisation. In \cite{UrbanPittsGabbay04} we gave a rather clunky proof
  for the property that the equivalence relation \isa{{\isaliteral{5C3C617070726F783E}{\isasymapprox}}} is
  transitive. This proof has been slightly simplified in \cite{FernandezGabbay07}.  The main
  purpose of this paper is to further simplify this proof. The idea behind the
  simplification is taken from the work of Kumar and Norrish who formalised
  nominal unification in the HOL4 theorem prover \cite{KumarNorrish10}, but did
  not report about their simplification in print. After describing the simpler
  proof in detail, we sketch the nominal unification algorithm and outline
  some results obtained about nominal unification.%
\end{isamarkuptext}%
\isamarkuptrue%
\isamarkupsection{Equality and Freshness\label{mainsec}%
}
\isamarkuptrue%
\begin{isamarkuptext}%
Two central notions in nominal unification are names, which are called \emph{atoms}, and 
  \emph{permutations} of atoms.
  We assume in this paper that there is a countably infinite set of atoms and represent permutations
  as finite lists of pairs of atoms. The elements of these lists are called \emph{swappings}.
  We therefore write permutations as \isa{{\isaliteral{28}{\isacharparenleft}}a\isaliteral{5C3C5E697375623E}{}\isactrlisub {\isadigit{1}}\ b\isaliteral{5C3C5E697375623E}{}\isactrlisub {\isadigit{1}}{\isaliteral{29}{\isacharparenright}}\ {\isaliteral{28}{\isacharparenleft}}a\isaliteral{5C3C5E697375623E}{}\isactrlisub {\isadigit{2}}\ b\isaliteral{5C3C5E697375623E}{}\isactrlisub {\isadigit{2}}{\isaliteral{29}{\isacharparenright}}\ {\isaliteral{5C3C646F74733E}{\isasymdots}}\ {\isaliteral{28}{\isacharparenleft}}a\isaliteral{5C3C5E697375623E}{}\isactrlisub n\ b\isaliteral{5C3C5E697375623E}{}\isactrlisub n{\isaliteral{29}{\isacharparenright}}}; the empty list
  \isa{{\isaliteral{5B}{\isacharbrackleft}}{\isaliteral{5D}{\isacharbrackright}}} stands for the identity permutation. A permutation \isa{{\isaliteral{5C3C70693E}{\isasympi}}} {\it
  acting} on an atom \isa{a} is defined as

  \[
  \begin{array}{rlcrcl}
    \isa{{\isaliteral{5C3C70693E}{\isasympi}}\ {\isaliteral{5C3C62756C6C65743E}{\isasymbullet}}\ a} & \dn & \isa{a}\qquad &
    \isa{{\isaliteral{28}{\isacharparenleft}}a\isaliteral{5C3C5E697375623E}{}\isactrlisub {\isadigit{1}}\ a\isaliteral{5C3C5E697375623E}{}\isactrlisub {\isadigit{2}}{\isaliteral{29}{\isacharparenright}}{\isaliteral{3A}{\isacharcolon}}{\isaliteral{3A}{\isacharcolon}}{\isaliteral{5C3C70693E}{\isasympi}}\ {\isaliteral{5C3C62756C6C65743E}{\isasymbullet}}\ a} & \dn &
    \left\{\begin{array}{ll}
        \isa{a\isaliteral{5C3C5E697375623E}{}\isactrlisub {\isadigit{2}}} & \text{if\;}\isa{{\isaliteral{5C3C70693E}{\isasympi}}\ {\isaliteral{5C3C62756C6C65743E}{\isasymbullet}}\ a\ {\isaliteral{3D}{\isacharequal}}\ a\isaliteral{5C3C5E697375623E}{}\isactrlisub {\isadigit{1}}}\\
        \isa{a\isaliteral{5C3C5E697375623E}{}\isactrlisub {\isadigit{1}}} & \text{if\;}\isa{{\isaliteral{5C3C70693E}{\isasympi}}\ {\isaliteral{5C3C62756C6C65743E}{\isasymbullet}}\ a\ {\isaliteral{3D}{\isacharequal}}\ a\isaliteral{5C3C5E697375623E}{}\isactrlisub {\isadigit{2}}}\\
        \isa{{\isaliteral{5C3C70693E}{\isasympi}}\ {\isaliteral{5C3C62756C6C65743E}{\isasymbullet}}\ a} & \text{otherwise}
      \end{array}\right.
  \end{array}\label{permatom}
  \]

  \noindent
  where  \isa{{\isaliteral{28}{\isacharparenleft}}a\isaliteral{5C3C5E697375623E}{}\isactrlisub {\isadigit{1}}\ a\isaliteral{5C3C5E697375623E}{}\isactrlisub {\isadigit{2}}{\isaliteral{29}{\isacharparenright}}{\isaliteral{3A}{\isacharcolon}}{\isaliteral{3A}{\isacharcolon}}{\isaliteral{5C3C70693E}{\isasympi}}} is the composition of a permutation
  followed by the swapping \isa{{\isaliteral{28}{\isacharparenleft}}a\isaliteral{5C3C5E697375623E}{}\isactrlisub {\isadigit{1}}\ a\isaliteral{5C3C5E697375623E}{}\isactrlisub {\isadigit{2}}{\isaliteral{29}{\isacharparenright}}}.  The composition of \isa{{\isaliteral{5C3C70693E}{\isasympi}}}
  followed by another permutation \isa{{\isaliteral{5C3C70693E}{\isasympi}}{\isaliteral{27}{\isacharprime}}} is given by list-concatenation,
  written as \isa{{\isaliteral{5C3C70693E}{\isasympi}}{\isaliteral{27}{\isacharprime}}\ {\isaliteral{40}{\isacharat}}\ {\isaliteral{5C3C70693E}{\isasympi}}}, and the inverse of a permutation is given by
  list reversal, written as \isa{{\isaliteral{5C3C70693E}{\isasympi}}\isaliteral{5C3C5E627375703E}{}\isactrlbsup {\isaliteral{2D}{\isacharminus}}{\isadigit{1}}\isaliteral{5C3C5E657375703E}{}\isactrlesup }.  

  The advantage of our representation of permutations-as-lists-of-swappings is
  that we can easily calculate the composition and the inverse of
  permutations, which are basic operations in the nominal unification
  algorithm. However, the list representation does not give unique
  representatives for permutations (for example \isa{{\isaliteral{28}{\isacharparenleft}}a\ a{\isaliteral{29}{\isacharparenright}}\ {\isaliteral{5C3C6E6F7465713E}{\isasymnoteq}}\ {\isaliteral{5B}{\isacharbrackleft}}{\isaliteral{5D}{\isacharbrackright}}}). This is 
  is different from the usual representation of permutations given for example in~\cite{HuffmanUrban10}.
  There permutations are (unique) bijective functions from atoms to atoms. For permutations-as-lists 
  we can define the \emph{disagreement set} between two permutations as the set of atoms 
  given by

  \begin{center}
  \isa{ds\ {\isaliteral{5C3C70693E}{\isasympi}}\ {\isaliteral{5C3C70693E}{\isasympi}}{\isaliteral{27}{\isacharprime}}\ {\isaliteral{5C3C65717569763E}{\isasymequiv}}\ {\isaliteral{7B}{\isacharbraceleft}}a\ {\isaliteral{7C}{\isacharbar}}\ {\isaliteral{5C3C70693E}{\isasympi}}\ {\isaliteral{5C3C62756C6C65743E}{\isasymbullet}}\ a\ {\isaliteral{5C3C6E6F7465713E}{\isasymnoteq}}\ {\isaliteral{5C3C70693E}{\isasympi}}{\isaliteral{27}{\isacharprime}}\ {\isaliteral{5C3C62756C6C65743E}{\isasymbullet}}\ a{\isaliteral{7D}{\isacharbraceright}}}
  \end{center}

  \noindent
  and then regard two permutations as equal provided their disagreement set is empty.
  However, we do not explicitly equate permutations.

  The purpose of unification is to make terms equal by substituting terms for variables. 
  The paper \cite{UrbanPittsGabbay04} defines \emph{nominal terms} with the following grammar:

  \begin{center}
  \begin{tabular}{rcll}
  \isa{trm} & \isa{{\isaliteral{3A}{\isacharcolon}}{\isaliteral{3A}{\isacharcolon}}{\isaliteral{3D}{\isacharequal}}} & \isa{{\isaliteral{5C3C6C616E676C653E}{\isasymlangle}}{\isaliteral{5C3C72616E676C653E}{\isasymrangle}}}       & Units\\
                & \isa{{\isaliteral{7C}{\isacharbar}}}   & \isa{{\isaliteral{5C3C6C616E676C653E}{\isasymlangle}}t\isaliteral{5C3C5E697375623E}{}\isactrlisub {\isadigit{1}}{\isaliteral{2C}{\isacharcomma}}\ t\isaliteral{5C3C5E697375623E}{}\isactrlisub {\isadigit{2}}{\isaliteral{5C3C72616E676C653E}{\isasymrangle}}} & Pairs\\
                & \isa{{\isaliteral{7C}{\isacharbar}}}   & \isa{f\ t}   & Function Symbols\\
                & \isa{{\isaliteral{7C}{\isacharbar}}}   & \isa{a}     & Atoms\\
                & \isa{{\isaliteral{7C}{\isacharbar}}}   & \isa{a{\isaliteral{2E}{\isachardot}}t}   & Abstractions\\
                & \isa{{\isaliteral{7C}{\isacharbar}}}   & \isa{{\isaliteral{5C3C70693E}{\isasympi}}{\isaliteral{5C3C63646F743E}{\isasymcdot}}X}   & Suspensions\\
  \end{tabular}
  \end{center}

  \noindent
  In order to slightly simplify the formal reasoning in the Isabelle/HOL
  theorem prover, the function symbols only take a single argument (instead of
  the usual list of arguments). Functions symbols with multiple arguments need
  to be encoded with pairs. An important point to note is that atoms,
  written \isa{a{\isaliteral{2C}{\isacharcomma}}\ b{\isaliteral{2C}{\isacharcomma}}\ c{\isaliteral{2C}{\isacharcomma}}\ {\isaliteral{5C3C646F74733E}{\isasymdots}}}, are distinct from variables, written \isa{X{\isaliteral{2C}{\isacharcomma}}\ Y{\isaliteral{2C}{\isacharcomma}}\ Z{\isaliteral{2C}{\isacharcomma}}\ {\isaliteral{5C3C646F74733E}{\isasymdots}}}, and only variables can potentially be substituted during nominal
  unification (a definition of substitution will be given shortly). As mentioned in the
  Introduction,
  variables in general need to be considered together with
  permutations---therefore suspensions are pairs consisting of a permutation and
  a variable. The reason for this definition is that variables stand for unknown 
  terms, and a permutation applied to a term must be 
  ``suspended'' in front of all unknowns in order to
  keep it for the case when any of the unknowns is substituted with a term.

  Another important point to note is that, although there are abstraction terms,
  nominal terms are first-order terms: there is no implicit 
  quotienting modulo renaming of bound names. For example the abstractions
  \isa{a{\isaliteral{2E}{\isachardot}}t} and \isa{b{\isaliteral{2E}{\isachardot}}s} are \emph{not} equal unless \isa{a\ {\isaliteral{3D}{\isacharequal}}\ b} and \isa{t\ {\isaliteral{3D}{\isacharequal}}\ s}.
  This has the advantage that nominal terms can be implemented as a simple
  datatype  in programming languages such as ML and also in the theorem prover Isabelle/HOL.
  In \cite{UrbanPittsGabbay04} a notion of \emph{equality} and \emph{freshness} 
  for nominal terms is defined by two inductive predicates whose rules are shown in 
  Figure~\ref{judgements}. This inductive definition uses freshness environments, written 
  \isa{{\isaliteral{5C3C6E61626C613E}{\isasymnabla}}}, which are sets of atom-and-variable pairs. We often write such environments as
  \isa{{\isaliteral{7B}{\isacharbraceleft}}a\isaliteral{5C3C5E697375623E}{}\isactrlisub {\isadigit{1}}\ {\isaliteral{23}{\isacharhash}}\ X\isaliteral{5C3C5E697375623E}{}\isactrlisub {\isadigit{1}}{\isaliteral{2C}{\isacharcomma}}\ {\isaliteral{5C3C646F74733E}{\isasymdots}}{\isaliteral{2C}{\isacharcomma}}\ a\isaliteral{5C3C5E697375623E}{}\isactrlisub n\ {\isaliteral{23}{\isacharhash}}\ X\isaliteral{5C3C5E697375623E}{}\isactrlisub n{\isaliteral{7D}{\isacharbraceright}}}. Rule (\isa{{\isaliteral{5C3C617070726F783E}{\isasymapprox}}}-abstraction$_2$) includes the operation of 
  applying a permutation to a nominal term, which can be recursively defined as

  \begin{center}
  \begin{tabular}{rcl}
  \isa{{\isaliteral{5C3C70693E}{\isasympi}}\ {\isaliteral{5C3C62756C6C65743E}{\isasymbullet}}\ {\isaliteral{28}{\isacharparenleft}}{\isaliteral{5C3C6C616E676C653E}{\isasymlangle}}{\isaliteral{5C3C72616E676C653E}{\isasymrangle}}{\isaliteral{29}{\isacharparenright}}} & \isa{{\isaliteral{5C3C65717569763E}{\isasymequiv}}} & \isa{{\isaliteral{5C3C6C616E676C653E}{\isasymlangle}}{\isaliteral{5C3C72616E676C653E}{\isasymrangle}}}\\
  \isa{{\isaliteral{5C3C70693E}{\isasympi}}\ {\isaliteral{5C3C62756C6C65743E}{\isasymbullet}}\ {\isaliteral{28}{\isacharparenleft}}{\isaliteral{5C3C6C616E676C653E}{\isasymlangle}}t\isaliteral{5C3C5E697375623E}{}\isactrlisub {\isadigit{1}}{\isaliteral{2C}{\isacharcomma}}\ t\isaliteral{5C3C5E697375623E}{}\isactrlisub {\isadigit{2}}{\isaliteral{5C3C72616E676C653E}{\isasymrangle}}{\isaliteral{29}{\isacharparenright}}} & \isa{{\isaliteral{5C3C65717569763E}{\isasymequiv}}} & \isa{{\isaliteral{5C3C6C616E676C653E}{\isasymlangle}}{\isaliteral{5C3C70693E}{\isasympi}}\ {\isaliteral{5C3C62756C6C65743E}{\isasymbullet}}\ t\isaliteral{5C3C5E697375623E}{}\isactrlisub {\isadigit{1}}{\isaliteral{2C}{\isacharcomma}}\ {\isaliteral{5C3C70693E}{\isasympi}}\ {\isaliteral{5C3C62756C6C65743E}{\isasymbullet}}\ t\isaliteral{5C3C5E697375623E}{}\isactrlisub {\isadigit{2}}{\isaliteral{5C3C72616E676C653E}{\isasymrangle}}}\\
  \isa{{\isaliteral{5C3C70693E}{\isasympi}}\ {\isaliteral{5C3C62756C6C65743E}{\isasymbullet}}\ {\isaliteral{28}{\isacharparenleft}}F\ t{\isaliteral{29}{\isacharparenright}}} & \isa{{\isaliteral{5C3C65717569763E}{\isasymequiv}}} & \isa{F\ {\isaliteral{28}{\isacharparenleft}}{\isaliteral{5C3C70693E}{\isasympi}}\ {\isaliteral{5C3C62756C6C65743E}{\isasymbullet}}\ t{\isaliteral{29}{\isacharparenright}}}\\
  \isa{{\isaliteral{5C3C70693E}{\isasympi}}\ {\isaliteral{5C3C62756C6C65743E}{\isasymbullet}}\ {\isaliteral{28}{\isacharparenleft}}{\isaliteral{5C3C70693E}{\isasympi}}{\isaliteral{27}{\isacharprime}}{\isaliteral{5C3C63646F743E}{\isasymcdot}}X{\isaliteral{29}{\isacharparenright}}} & \isa{{\isaliteral{5C3C65717569763E}{\isasymequiv}}} & \isa{{\isaliteral{28}{\isacharparenleft}}{\isaliteral{5C3C70693E}{\isasympi}}\ \ {\isaliteral{40}{\isacharat}}\ {\isaliteral{5C3C70693E}{\isasympi}}{\isaliteral{27}{\isacharprime}}{\isaliteral{29}{\isacharparenright}}{\isaliteral{5C3C63646F743E}{\isasymcdot}}X}\\
  \isa{{\isaliteral{5C3C70693E}{\isasympi}}\ {\isaliteral{5C3C62756C6C65743E}{\isasymbullet}}\ {\isaliteral{28}{\isacharparenleft}}a{\isaliteral{2E}{\isachardot}}\ t{\isaliteral{29}{\isacharparenright}}} & \isa{{\isaliteral{5C3C65717569763E}{\isasymequiv}}} & \isa{{\isaliteral{28}{\isacharparenleft}}{\isaliteral{5C3C70693E}{\isasympi}}\ {\isaliteral{5C3C62756C6C65743E}{\isasymbullet}}\ a{\isaliteral{29}{\isacharparenright}}{\isaliteral{2E}{\isachardot}}\ {\isaliteral{28}{\isacharparenleft}}{\isaliteral{5C3C70693E}{\isasympi}}\ {\isaliteral{5C3C62756C6C65743E}{\isasymbullet}}\ t{\isaliteral{29}{\isacharparenright}}}\\  
  \end{tabular}
  \end{center}
  
  \noindent
  where the clause for atoms is given in \eqref{permatom}. Because we suspend permutations
  in front of variables (see penultimate clause), it will in general be the case that
  
  \begin{equation}\label{nothold}
  \isa{{\isaliteral{5C3C70693E}{\isasympi}}\ {\isaliteral{5C3C62756C6C65743E}{\isasymbullet}}\ t\ {\isaliteral{5C3C6E6F7465713E}{\isasymnoteq}}\ {\isaliteral{5C3C70693E}{\isasympi}}{\isaliteral{27}{\isacharprime}}\ {\isaliteral{5C3C62756C6C65743E}{\isasymbullet}}\ t}
  \end{equation}

  \noindent 
  even if the disagreement set of \isa{{\isaliteral{5C3C70693E}{\isasympi}}} and \isa{{\isaliteral{5C3C70693E}{\isasympi}}{\isaliteral{27}{\isacharprime}}} is empty. Note
  that permutations acting on abstractions will permute both, the ``binder'' \isa{a}
  and the ``body'' \isa{t}.

  \begin{figure}[t]
  \begin{center}
  \begin{tabular}{c}

  \isa{\mbox{}\inferrule{\mbox{}}{\mbox{{\isaliteral{5C3C6E61626C613E}{\isasymnabla}}\ {\isaliteral{5C3C7475726E7374696C653E}{\isasymturnstile}}\ a\ {\isaliteral{23}{\isacharhash}}\ {\isaliteral{5C3C6C616E676C653E}{\isasymlangle}}{\isaliteral{5C3C72616E676C653E}{\isasymrangle}}}}}(\isa{{\isaliteral{23}{\isacharhash}}}-unit)\hspace{5mm}
  \isa{\mbox{}\inferrule{\mbox{{\isaliteral{5C3C6E61626C613E}{\isasymnabla}}\ {\isaliteral{5C3C7475726E7374696C653E}{\isasymturnstile}}\ a\ {\isaliteral{23}{\isacharhash}}\ t\isaliteral{5C3C5E697375623E}{}\isactrlisub {\isadigit{1}}}\\\ \mbox{{\isaliteral{5C3C6E61626C613E}{\isasymnabla}}\ {\isaliteral{5C3C7475726E7374696C653E}{\isasymturnstile}}\ a\ {\isaliteral{23}{\isacharhash}}\ t\isaliteral{5C3C5E697375623E}{}\isactrlisub {\isadigit{2}}}}{\mbox{{\isaliteral{5C3C6E61626C613E}{\isasymnabla}}\ {\isaliteral{5C3C7475726E7374696C653E}{\isasymturnstile}}\ a\ {\isaliteral{23}{\isacharhash}}\ {\isaliteral{5C3C6C616E676C653E}{\isasymlangle}}t\isaliteral{5C3C5E697375623E}{}\isactrlisub {\isadigit{1}}{\isaliteral{2C}{\isacharcomma}}\ t\isaliteral{5C3C5E697375623E}{}\isactrlisub {\isadigit{2}}{\isaliteral{5C3C72616E676C653E}{\isasymrangle}}}}}(\isa{{\isaliteral{23}{\isacharhash}}}-pair)
  \hspace{5mm}
  \isa{\mbox{}\inferrule{\mbox{{\isaliteral{5C3C6E61626C613E}{\isasymnabla}}\ {\isaliteral{5C3C7475726E7374696C653E}{\isasymturnstile}}\ a\ {\isaliteral{23}{\isacharhash}}\ t}}{\mbox{{\isaliteral{5C3C6E61626C613E}{\isasymnabla}}\ {\isaliteral{5C3C7475726E7374696C653E}{\isasymturnstile}}\ a\ {\isaliteral{23}{\isacharhash}}\ f\ t}}}(\isa{{\isaliteral{23}{\isacharhash}}}-function symbol)\medskip\\

  \isa{\mbox{}\inferrule{\mbox{}}{\mbox{{\isaliteral{5C3C6E61626C613E}{\isasymnabla}}\ {\isaliteral{5C3C7475726E7374696C653E}{\isasymturnstile}}\ a\ {\isaliteral{23}{\isacharhash}}\ a{\isaliteral{2E}{\isachardot}}t}}}(\isa{{\isaliteral{23}{\isacharhash}}}-abstraction$_1$)\hspace{5mm}
  \isa{\mbox{}\inferrule{\mbox{{\isaliteral{5C3C6E61626C613E}{\isasymnabla}}\ {\isaliteral{5C3C7475726E7374696C653E}{\isasymturnstile}}\ a\ {\isaliteral{23}{\isacharhash}}\ t}\\\ \mbox{a\ {\isaliteral{5C3C6E6F7465713E}{\isasymnoteq}}\ b}}{\mbox{{\isaliteral{5C3C6E61626C613E}{\isasymnabla}}\ {\isaliteral{5C3C7475726E7374696C653E}{\isasymturnstile}}\ a\ {\isaliteral{23}{\isacharhash}}\ b{\isaliteral{2E}{\isachardot}}t}}}(\isa{{\isaliteral{23}{\isacharhash}}}-abstraction$_2$)\medskip\\

  \isa{\mbox{}\inferrule{\mbox{a\ {\isaliteral{5C3C6E6F7465713E}{\isasymnoteq}}\ b}}{\mbox{{\isaliteral{5C3C6E61626C613E}{\isasymnabla}}\ {\isaliteral{5C3C7475726E7374696C653E}{\isasymturnstile}}\ a\ {\isaliteral{23}{\isacharhash}}\ b}}}(\isa{{\isaliteral{23}{\isacharhash}}}-atom)\hspace{5mm}
  \isa{\mbox{}\inferrule{\mbox{{\isaliteral{28}{\isacharparenleft}}{\isaliteral{5C3C70693E}{\isasympi}}\isaliteral{5C3C5E627375703E}{}\isactrlbsup {\isaliteral{2D}{\isacharminus}}{\isadigit{1}}\isaliteral{5C3C5E657375703E}{}\isactrlesup \ {\isaliteral{5C3C62756C6C65743E}{\isasymbullet}}\ a{\isaliteral{2C}{\isacharcomma}}\ X{\isaliteral{29}{\isacharparenright}}\ {\isaliteral{5C3C696E3E}{\isasymin}}\ {\isaliteral{5C3C6E61626C613E}{\isasymnabla}}}}{\mbox{{\isaliteral{5C3C6E61626C613E}{\isasymnabla}}\ {\isaliteral{5C3C7475726E7374696C653E}{\isasymturnstile}}\ a\ {\isaliteral{23}{\isacharhash}}\ {\isaliteral{5C3C70693E}{\isasympi}}{\isaliteral{5C3C63646F743E}{\isasymcdot}}X}}}(\isa{{\isaliteral{23}{\isacharhash}}}-suspension)\bigskip\\

  \isa{\mbox{}\inferrule{\mbox{}}{\mbox{\ {\isaliteral{5C3C6E61626C613E}{\isasymnabla}}\ {\isaliteral{5C3C7475726E7374696C653E}{\isasymturnstile}}\ {\isaliteral{5C3C6C616E676C653E}{\isasymlangle}}{\isaliteral{5C3C72616E676C653E}{\isasymrangle}}\ {\isaliteral{5C3C617070726F783E}{\isasymapprox}}\ {\isaliteral{5C3C6C616E676C653E}{\isasymlangle}}{\isaliteral{5C3C72616E676C653E}{\isasymrangle}}}}}(\isa{{\isaliteral{5C3C617070726F783E}{\isasymapprox}}}-unit)\hspace{5mm}
  \isa{\mbox{}\inferrule{\mbox{\ {\isaliteral{5C3C6E61626C613E}{\isasymnabla}}\ {\isaliteral{5C3C7475726E7374696C653E}{\isasymturnstile}}\ t\isaliteral{5C3C5E697375623E}{}\isactrlisub {\isadigit{1}}\ {\isaliteral{5C3C617070726F783E}{\isasymapprox}}\ t\isaliteral{5C3C5E697375623E}{}\isactrlisub {\isadigit{2}}}\\\ \mbox{\ {\isaliteral{5C3C6E61626C613E}{\isasymnabla}}\ {\isaliteral{5C3C7475726E7374696C653E}{\isasymturnstile}}\ s\isaliteral{5C3C5E697375623E}{}\isactrlisub {\isadigit{1}}\ {\isaliteral{5C3C617070726F783E}{\isasymapprox}}\ s\isaliteral{5C3C5E697375623E}{}\isactrlisub {\isadigit{2}}}}{\mbox{\ {\isaliteral{5C3C6E61626C613E}{\isasymnabla}}\ {\isaliteral{5C3C7475726E7374696C653E}{\isasymturnstile}}\ {\isaliteral{5C3C6C616E676C653E}{\isasymlangle}}t\isaliteral{5C3C5E697375623E}{}\isactrlisub {\isadigit{1}}{\isaliteral{2C}{\isacharcomma}}\ s\isaliteral{5C3C5E697375623E}{}\isactrlisub {\isadigit{1}}{\isaliteral{5C3C72616E676C653E}{\isasymrangle}}\ {\isaliteral{5C3C617070726F783E}{\isasymapprox}}\ {\isaliteral{5C3C6C616E676C653E}{\isasymlangle}}t\isaliteral{5C3C5E697375623E}{}\isactrlisub {\isadigit{2}}{\isaliteral{2C}{\isacharcomma}}\ s\isaliteral{5C3C5E697375623E}{}\isactrlisub {\isadigit{2}}{\isaliteral{5C3C72616E676C653E}{\isasymrangle}}}}}(\isa{{\isaliteral{5C3C617070726F783E}{\isasymapprox}}}-pair)\hspace{5mm}
  \isa{\mbox{}\inferrule{\mbox{\ {\isaliteral{5C3C6E61626C613E}{\isasymnabla}}\ {\isaliteral{5C3C7475726E7374696C653E}{\isasymturnstile}}\ t\isaliteral{5C3C5E697375623E}{}\isactrlisub {\isadigit{1}}\ {\isaliteral{5C3C617070726F783E}{\isasymapprox}}\ t\isaliteral{5C3C5E697375623E}{}\isactrlisub {\isadigit{2}}}}{\mbox{\ {\isaliteral{5C3C6E61626C613E}{\isasymnabla}}\ {\isaliteral{5C3C7475726E7374696C653E}{\isasymturnstile}}\ f\ t\isaliteral{5C3C5E697375623E}{}\isactrlisub {\isadigit{1}}\ {\isaliteral{5C3C617070726F783E}{\isasymapprox}}\ f\ t\isaliteral{5C3C5E697375623E}{}\isactrlisub {\isadigit{2}}}}}%
  (\isa{{\isaliteral{5C3C617070726F783E}{\isasymapprox}}}-function symbol)
  \medskip\\

  \isa{\mbox{}\inferrule{\mbox{\ {\isaliteral{5C3C6E61626C613E}{\isasymnabla}}\ {\isaliteral{5C3C7475726E7374696C653E}{\isasymturnstile}}\ t\isaliteral{5C3C5E697375623E}{}\isactrlisub {\isadigit{1}}\ {\isaliteral{5C3C617070726F783E}{\isasymapprox}}\ t\isaliteral{5C3C5E697375623E}{}\isactrlisub {\isadigit{2}}}}{\mbox{\ {\isaliteral{5C3C6E61626C613E}{\isasymnabla}}\ {\isaliteral{5C3C7475726E7374696C653E}{\isasymturnstile}}\ a{\isaliteral{2E}{\isachardot}}t\isaliteral{5C3C5E697375623E}{}\isactrlisub {\isadigit{1}}\ {\isaliteral{5C3C617070726F783E}{\isasymapprox}}\ a{\isaliteral{2E}{\isachardot}}t\isaliteral{5C3C5E697375623E}{}\isactrlisub {\isadigit{2}}}}}%
  (\isa{{\isaliteral{5C3C617070726F783E}{\isasymapprox}}}-abstraction$_1$)\hspace{5mm}
  \isa{\mbox{}\inferrule{\mbox{a\ {\isaliteral{5C3C6E6F7465713E}{\isasymnoteq}}\ b}\\\ \mbox{{\isaliteral{5C3C6E61626C613E}{\isasymnabla}}\ {\isaliteral{5C3C7475726E7374696C653E}{\isasymturnstile}}\ a\ {\isaliteral{23}{\isacharhash}}\ t\isaliteral{5C3C5E697375623E}{}\isactrlisub {\isadigit{2}}}\\\ \mbox{\ {\isaliteral{5C3C6E61626C613E}{\isasymnabla}}\ {\isaliteral{5C3C7475726E7374696C653E}{\isasymturnstile}}\ t\isaliteral{5C3C5E697375623E}{}\isactrlisub {\isadigit{1}}\ {\isaliteral{5C3C617070726F783E}{\isasymapprox}}\ {\isaliteral{28}{\isacharparenleft}}a\ b{\isaliteral{29}{\isacharparenright}}\ {\isaliteral{5C3C63646F743E}{\isasymcdot}}\ t\isaliteral{5C3C5E697375623E}{}\isactrlisub {\isadigit{2}}}}{\mbox{\ {\isaliteral{5C3C6E61626C613E}{\isasymnabla}}\ {\isaliteral{5C3C7475726E7374696C653E}{\isasymturnstile}}\ a{\isaliteral{2E}{\isachardot}}t\isaliteral{5C3C5E697375623E}{}\isactrlisub {\isadigit{1}}\ {\isaliteral{5C3C617070726F783E}{\isasymapprox}}\ b{\isaliteral{2E}{\isachardot}}t\isaliteral{5C3C5E697375623E}{}\isactrlisub {\isadigit{2}}}}}%
  (\isa{{\isaliteral{5C3C617070726F783E}{\isasymapprox}}}-abstraction$_2$)\medskip\\

  \isa{\mbox{}\inferrule{\mbox{}}{\mbox{\ {\isaliteral{5C3C6E61626C613E}{\isasymnabla}}\ {\isaliteral{5C3C7475726E7374696C653E}{\isasymturnstile}}\ a\ {\isaliteral{5C3C617070726F783E}{\isasymapprox}}\ a}}}(\isa{{\isaliteral{5C3C617070726F783E}{\isasymapprox}}}-atom)\hspace{5mm}
  \isa{\mbox{}\inferrule{\mbox{{\isaliteral{5C3C666F72616C6C3E}{\isasymforall}}c{\isaliteral{5C3C696E3E}{\isasymin}}ds\ {\isaliteral{5C3C70693E}{\isasympi}}\ {\isaliteral{5C3C70693E}{\isasympi}}{\isaliteral{27}{\isacharprime}}{\isaliteral{2E}{\isachardot}}\ {\isaliteral{28}{\isacharparenleft}}c{\isaliteral{2C}{\isacharcomma}}\ X{\isaliteral{29}{\isacharparenright}}\ {\isaliteral{5C3C696E3E}{\isasymin}}\ {\isaliteral{5C3C6E61626C613E}{\isasymnabla}}}}{\mbox{\ {\isaliteral{5C3C6E61626C613E}{\isasymnabla}}\ {\isaliteral{5C3C7475726E7374696C653E}{\isasymturnstile}}\ {\isaliteral{5C3C70693E}{\isasympi}}{\isaliteral{5C3C63646F743E}{\isasymcdot}}X\ {\isaliteral{5C3C617070726F783E}{\isasymapprox}}\ {\isaliteral{5C3C70693E}{\isasympi}}{\isaliteral{27}{\isacharprime}}{\isaliteral{5C3C63646F743E}{\isasymcdot}}X}}}(\isa{{\isaliteral{5C3C617070726F783E}{\isasymapprox}}}-suspension)  
  \end{tabular}
  \end{center}

  \caption{Inductive definitions for freshness and equality of nominal terms.\label{judgements}}
  \end{figure}

  In order to show the correctness of the nominal unification algorithm in \cite{UrbanPittsGabbay04}, 
  one first needs to establish that \isa{{\isaliteral{5C3C617070726F783E}{\isasymapprox}}} is an equivalence relation in the sense of

  \begin{center}
  \begin{tabular}{rl@ {\hspace{10mm}}l}
  \isa{{\isaliteral{28}{\isacharparenleft}}i{\isaliteral{29}{\isacharparenright}}}   & \isa{{\isaliteral{5C3C6E61626C613E}{\isasymnabla}}\ {\isaliteral{5C3C7475726E7374696C653E}{\isasymturnstile}}\ t\ {\isaliteral{5C3C617070726F783E}{\isasymapprox}}\ t} & (reflexivity)\\ 
  \isa{{\isaliteral{28}{\isacharparenleft}}ii{\isaliteral{29}{\isacharparenright}}}  & \isa{{\isaliteral{5C3C6E61626C613E}{\isasymnabla}}\ {\isaliteral{5C3C7475726E7374696C653E}{\isasymturnstile}}\ t\isaliteral{5C3C5E697375623E}{}\isactrlisub {\isadigit{1}}\ {\isaliteral{5C3C617070726F783E}{\isasymapprox}}\ t\isaliteral{5C3C5E697375623E}{}\isactrlisub {\isadigit{2}}} implies \isa{{\isaliteral{5C3C6E61626C613E}{\isasymnabla}}\ {\isaliteral{5C3C7475726E7374696C653E}{\isasymturnstile}}\ t\isaliteral{5C3C5E697375623E}{}\isactrlisub {\isadigit{2}}\ {\isaliteral{5C3C617070726F783E}{\isasymapprox}}\ t\isaliteral{5C3C5E697375623E}{}\isactrlisub {\isadigit{1}}} & (symmetry)\\ 
  \isa{{\isaliteral{28}{\isacharparenleft}}iii{\isaliteral{29}{\isacharparenright}}} & \isa{{\isaliteral{5C3C6E61626C613E}{\isasymnabla}}\ {\isaliteral{5C3C7475726E7374696C653E}{\isasymturnstile}}\ t\isaliteral{5C3C5E697375623E}{}\isactrlisub {\isadigit{1}}\ {\isaliteral{5C3C617070726F783E}{\isasymapprox}}\ t\isaliteral{5C3C5E697375623E}{}\isactrlisub {\isadigit{2}}} and \isa{{\isaliteral{5C3C6E61626C613E}{\isasymnabla}}\ {\isaliteral{5C3C7475726E7374696C653E}{\isasymturnstile}}\ t\isaliteral{5C3C5E697375623E}{}\isactrlisub {\isadigit{2}}\ {\isaliteral{5C3C617070726F783E}{\isasymapprox}}\ t\isaliteral{5C3C5E697375623E}{}\isactrlisub {\isadigit{3}}} imply \isa{{\isaliteral{5C3C6E61626C613E}{\isasymnabla}}\ {\isaliteral{5C3C7475726E7374696C653E}{\isasymturnstile}}\ t\isaliteral{5C3C5E697375623E}{}\isactrlisub {\isadigit{1}}\ {\isaliteral{5C3C617070726F783E}{\isasymapprox}}\ t\isaliteral{5C3C5E697375623E}{}\isactrlisub {\isadigit{3}}}
  & (transitivity)\\ 
  \end{tabular}
  \end{center}

  \noindent
  The first property can be proved by a routine induction over the structure of \isa{t}. 
  Given the ``unsymmetric'' formulation of the (\isa{{\isaliteral{5C3C617070726F783E}{\isasymapprox}}}-abstraction$_2$) rule, 
  the fact that \isa{{\isaliteral{5C3C617070726F783E}{\isasymapprox}}} is symmetric is at first glance surprising. Furthermore, a 
  direct proof by induction over the rules seems tricky, since in 
  the (\isa{{\isaliteral{5C3C617070726F783E}{\isasymapprox}}}-abstraction$_2$) case one needs to infer 
  \isa{{\isaliteral{5C3C6E61626C613E}{\isasymnabla}}\ {\isaliteral{5C3C7475726E7374696C653E}{\isasymturnstile}}\ t\isaliteral{5C3C5E697375623E}{}\isactrlisub {\isadigit{2}}\ {\isaliteral{5C3C617070726F783E}{\isasymapprox}}\ {\isaliteral{28}{\isacharparenleft}}b\ a{\isaliteral{29}{\isacharparenright}}\ {\isaliteral{5C3C62756C6C65743E}{\isasymbullet}}\ t\isaliteral{5C3C5E697375623E}{}\isactrlisub {\isadigit{1}}} from \isa{{\isaliteral{5C3C6E61626C613E}{\isasymnabla}}\ {\isaliteral{5C3C7475726E7374696C653E}{\isasymturnstile}}\ \ {\isaliteral{28}{\isacharparenleft}}a\ b{\isaliteral{29}{\isacharparenright}}\ {\isaliteral{5C3C62756C6C65743E}{\isasymbullet}}\ t\isaliteral{5C3C5E697375623E}{}\isactrlisub {\isadigit{2}}\ {\isaliteral{5C3C617070726F783E}{\isasymapprox}}\ t\isaliteral{5C3C5E697375623E}{}\isactrlisub {\isadigit{1}}}. This needs
  several supporting lemmas about freshness and equality, but ultimately requires
  that the transitivity property is proved first. Unfortunately, 
  a direct proof by rule-induction for transitivity seems even more difficult and 
  we did not manage to find one in \cite{UrbanPittsGabbay04}. Instead we resorted 
  to a clunky induction over the 
  size of terms (since size is preserved under permutations). To make matters
  worse, this induction over the
  size of terms needed to be
  loaded with two more properties in order to get 
  the induction through. The authors of \cite{FernandezGabbay07} managed to split 
  up this bulky induction, but still relied on an induction over the size of terms
  in their transitivity proof.

  The authors of \cite{KumarNorrish10} managed to do considerably better.
  They use a clever trick in their formalisation of nominal unification in HOL4 
  (their proof of equivalence is not shown in the paper).
  This trick yields a simpler and more direct proof for transitivity, than the ones given in 
  \cite{UrbanPittsGabbay04,FernandezGabbay07}. We shall below adapt the proof 
  by Kumar and Norrish to our  setting of (first-order) nominal 
  terms\footnote{Their formalisation in HOL4 introduces an
  indirection by using a quotient construction over nominal terms. This quotient
  construction does not translate into a simple datatype definition for 
  nominal terms.}.
  First we can establish the following property. 

  \begin{lemma}\label{fresheqvt}
  If \isa{{\isaliteral{5C3C6E61626C613E}{\isasymnabla}}\ {\isaliteral{5C3C7475726E7374696C653E}{\isasymturnstile}}\ a\ {\isaliteral{23}{\isacharhash}}\ t} then also \isa{{\isaliteral{5C3C6E61626C613E}{\isasymnabla}}\ {\isaliteral{5C3C7475726E7374696C653E}{\isasymturnstile}}\ {\isaliteral{28}{\isacharparenleft}}{\isaliteral{5C3C70693E}{\isasympi}}\ {\isaliteral{5C3C62756C6C65743E}{\isasymbullet}}\ a{\isaliteral{29}{\isacharparenright}}\ {\isaliteral{23}{\isacharhash}}\ {\isaliteral{28}{\isacharparenleft}}{\isaliteral{5C3C70693E}{\isasympi}}\ {\isaliteral{5C3C62756C6C65743E}{\isasymbullet}}\ t{\isaliteral{29}{\isacharparenright}}}, and vice versa.
  \end{lemma}

  \noindent
  The proof is by a routine induction on the structure of \isa{t} and we omit the details. 
  Following \cite{UrbanPittsGabbay04}
  we can next attempt to prove that freshness is preserved under equality (Lemma \ref{freshequ} below). 
  However here the trick from \cite{KumarNorrish10} already helps to simplify the reasoning. In 
  \cite{KumarNorrish10} the notion of \emph{weak equivalence}, 
  written as \isa{{\isaliteral{5C3C73696D3E}{\isasymsim}}}, is defined as follows

  \begin{center}
  \begin{tabular}{c}
  \isa{\mbox{}\inferrule{\mbox{}}{\mbox{{\isaliteral{5C3C6C616E676C653E}{\isasymlangle}}{\isaliteral{5C3C72616E676C653E}{\isasymrangle}}\ {\isaliteral{5C3C73696D3E}{\isasymsim}}\ {\isaliteral{5C3C6C616E676C653E}{\isasymlangle}}{\isaliteral{5C3C72616E676C653E}{\isasymrangle}}}}}\hspace{5mm}
  \isa{\mbox{}\inferrule{\mbox{}}{\mbox{a\ {\isaliteral{5C3C73696D3E}{\isasymsim}}\ a}}}\hspace{5mm}
  \isa{\mbox{}\inferrule{\mbox{t\ {\isaliteral{5C3C73696D3E}{\isasymsim}}\ t{\isaliteral{27}{\isacharprime}}}}{\mbox{f\ t\ {\isaliteral{5C3C73696D3E}{\isasymsim}}\ f\ t{\isaliteral{27}{\isacharprime}}}}}\medskip\\
  
  \isa{\mbox{}\inferrule{\mbox{t\isaliteral{5C3C5E697375623E}{}\isactrlisub {\isadigit{1}}\ {\isaliteral{5C3C73696D3E}{\isasymsim}}\ s\isaliteral{5C3C5E697375623E}{}\isactrlisub {\isadigit{1}}}\\\ \mbox{t\isaliteral{5C3C5E697375623E}{}\isactrlisub {\isadigit{2}}\ {\isaliteral{5C3C73696D3E}{\isasymsim}}\ s\isaliteral{5C3C5E697375623E}{}\isactrlisub {\isadigit{2}}}}{\mbox{{\isaliteral{5C3C6C616E676C653E}{\isasymlangle}}t\isaliteral{5C3C5E697375623E}{}\isactrlisub {\isadigit{1}}{\isaliteral{2C}{\isacharcomma}}\ t\isaliteral{5C3C5E697375623E}{}\isactrlisub {\isadigit{2}}{\isaliteral{5C3C72616E676C653E}{\isasymrangle}}\ {\isaliteral{5C3C73696D3E}{\isasymsim}}\ {\isaliteral{5C3C6C616E676C653E}{\isasymlangle}}s\isaliteral{5C3C5E697375623E}{}\isactrlisub {\isadigit{1}}{\isaliteral{2C}{\isacharcomma}}\ s\isaliteral{5C3C5E697375623E}{}\isactrlisub {\isadigit{2}}{\isaliteral{5C3C72616E676C653E}{\isasymrangle}}}}}\hspace{5mm}
  \isa{\mbox{}\inferrule{\mbox{t\ {\isaliteral{5C3C73696D3E}{\isasymsim}}\ t{\isaliteral{27}{\isacharprime}}}}{\mbox{a{\isaliteral{2E}{\isachardot}}t\ {\isaliteral{5C3C73696D3E}{\isasymsim}}\ a{\isaliteral{2E}{\isachardot}}t{\isaliteral{27}{\isacharprime}}}}}\hspace{5mm}
  \isa{\mbox{}\inferrule{\mbox{ds\ {\isaliteral{5C3C70693E}{\isasympi}}\ {\isaliteral{5C3C70693E}{\isasympi}}{\isaliteral{27}{\isacharprime}}\ {\isaliteral{3D}{\isacharequal}}\ {\isaliteral{5C3C656D7074797365743E}{\isasymemptyset}}}}{\mbox{{\isaliteral{5C3C70693E}{\isasympi}}{\isaliteral{5C3C63646F743E}{\isasymcdot}}X\ {\isaliteral{5C3C73696D3E}{\isasymsim}}\ {\isaliteral{5C3C70693E}{\isasympi}}{\isaliteral{27}{\isacharprime}}{\isaliteral{5C3C63646F743E}{\isasymcdot}}X}}}
  \end{tabular}
  \end{center}

  \noindent
  This equivalence is said to be weak because two terms can only differ in the permutations that
  are suspended in front of variables. Moreover, these permutations can only be equal (in the sense that is 
  their disagreement set must be empty). One advantage of this definition is that we can show 

  \begin{equation}\label{weakpi}
  \mbox{\isa{{\isaliteral{5C3C70693E}{\isasympi}}\ {\isaliteral{5C3C62756C6C65743E}{\isasymbullet}}\ t\ {\isaliteral{5C3C73696D3E}{\isasymsim}}\ {\isaliteral{5C3C70693E}{\isasympi}}{\isaliteral{27}{\isacharprime}}\ {\isaliteral{5C3C62756C6C65743E}{\isasymbullet}}\ t}~~provided~~\isa{ds\ {\isaliteral{5C3C70693E}{\isasympi}}\ {\isaliteral{5C3C70693E}{\isasympi}}{\isaliteral{27}{\isacharprime}}\ {\isaliteral{3D}{\isacharequal}}\ {\isaliteral{5C3C656D7074797365743E}{\isasymemptyset}}}}
  \end{equation}

  \noindent
  by an easy induction on \isa{t}. As noted in \eqref{nothold}, this property does 
  not hold when formulated with \isa{{\isaliteral{3D}{\isacharequal}}}. It is also straightforward to show that

  \begin{lemma}\label{weak}\mbox{}\\
  \isa{{\isaliteral{28}{\isacharparenleft}}i{\isaliteral{29}{\isacharparenright}}} \isa{{\normalsize{}If\,}\ \mbox{{\isaliteral{5C3C6E61626C613E}{\isasymnabla}}\ {\isaliteral{5C3C7475726E7374696C653E}{\isasymturnstile}}\ a\ {\isaliteral{23}{\isacharhash}}\ t\isaliteral{5C3C5E697375623E}{}\isactrlisub {\isadigit{1}}}\ {\normalsize \,and\,}\ \mbox{t\isaliteral{5C3C5E697375623E}{}\isactrlisub {\isadigit{1}}\ {\isaliteral{5C3C73696D3E}{\isasymsim}}\ t\isaliteral{5C3C5E697375623E}{}\isactrlisub {\isadigit{2}}}\ {\normalsize \,then\,}\ {\isaliteral{5C3C6E61626C613E}{\isasymnabla}}\ {\isaliteral{5C3C7475726E7374696C653E}{\isasymturnstile}}\ a\ {\isaliteral{23}{\isacharhash}}\ t\isaliteral{5C3C5E697375623E}{}\isactrlisub {\isadigit{2}}{\isaliteral{2E}{\isachardot}}}\\
  \isa{{\isaliteral{28}{\isacharparenleft}}ii{\isaliteral{29}{\isacharparenright}}} \isa{{\normalsize{}If\,}\ \mbox{\ {\isaliteral{5C3C6E61626C613E}{\isasymnabla}}\ {\isaliteral{5C3C7475726E7374696C653E}{\isasymturnstile}}\ t\isaliteral{5C3C5E697375623E}{}\isactrlisub {\isadigit{1}}\ {\isaliteral{5C3C617070726F783E}{\isasymapprox}}\ t\isaliteral{5C3C5E697375623E}{}\isactrlisub {\isadigit{2}}}\ {\normalsize \,and\,}\ \mbox{t\isaliteral{5C3C5E697375623E}{}\isactrlisub {\isadigit{2}}\ {\isaliteral{5C3C73696D3E}{\isasymsim}}\ t\isaliteral{5C3C5E697375623E}{}\isactrlisub {\isadigit{3}}}\ {\normalsize \,then\,}\ \ {\isaliteral{5C3C6E61626C613E}{\isasymnabla}}\ {\isaliteral{5C3C7475726E7374696C653E}{\isasymturnstile}}\ t\isaliteral{5C3C5E697375623E}{}\isactrlisub {\isadigit{1}}\ {\isaliteral{5C3C617070726F783E}{\isasymapprox}}\ t\isaliteral{5C3C5E697375623E}{}\isactrlisub {\isadigit{3}}{\isaliteral{2E}{\isachardot}}}
  \end{lemma}

  \noindent
  by induction over the relations \isa{{\isaliteral{5C3C73696D3E}{\isasymsim}}} and \isa{{\isaliteral{5C3C617070726F783E}{\isasymapprox}}}, respectively. The reason
  that these inductions go through with ease is that the relation \isa{{\isaliteral{5C3C73696D3E}{\isasymsim}}} 
  excludes the tricky cases where abstractions differ in their ``bound'' atoms. Using these two properties 
  together with \eqref{weakpi}, it is straightforward to establish:

  \begin{lemma}\label{freshequ}
  \isa{{\normalsize{}If\,}\ \mbox{\ {\isaliteral{5C3C6E61626C613E}{\isasymnabla}}\ {\isaliteral{5C3C7475726E7374696C653E}{\isasymturnstile}}\ t\isaliteral{5C3C5E697375623E}{}\isactrlisub {\isadigit{1}}\ {\isaliteral{5C3C617070726F783E}{\isasymapprox}}\ t\isaliteral{5C3C5E697375623E}{}\isactrlisub {\isadigit{2}}}\ {\normalsize \,and\,}\ \mbox{{\isaliteral{5C3C6E61626C613E}{\isasymnabla}}\ {\isaliteral{5C3C7475726E7374696C653E}{\isasymturnstile}}\ a\ {\isaliteral{23}{\isacharhash}}\ t\isaliteral{5C3C5E697375623E}{}\isactrlisub {\isadigit{1}}}\ {\normalsize \,then\,}\ {\isaliteral{5C3C6E61626C613E}{\isasymnabla}}\ {\isaliteral{5C3C7475726E7374696C653E}{\isasymturnstile}}\ a\ {\isaliteral{23}{\isacharhash}}\ t\isaliteral{5C3C5E697375623E}{}\isactrlisub {\isadigit{2}}{\isaliteral{2E}{\isachardot}}}
  \end{lemma}

  \begin{proof}
  By induction on the first judgement. The only interesting case is the rule 
  (\isa{{\isaliteral{5C3C617070726F783E}{\isasymapprox}}}-abstraction$_2$) where
  we need to establish \isa{{\isaliteral{5C3C6E61626C613E}{\isasymnabla}}\ {\isaliteral{5C3C7475726E7374696C653E}{\isasymturnstile}}\ a\ {\isaliteral{23}{\isacharhash}}\ d{\isaliteral{2E}{\isachardot}}t\isaliteral{5C3C5E697375623E}{}\isactrlisub {\isadigit{2}}} from the assumption \isa{{\isaliteral{28}{\isacharparenleft}}{\isaliteral{2A}{\isacharasterisk}}{\isaliteral{29}{\isacharparenright}}} 
  \isa{{\isaliteral{5C3C6E61626C613E}{\isasymnabla}}\ {\isaliteral{5C3C7475726E7374696C653E}{\isasymturnstile}}\ a\ {\isaliteral{23}{\isacharhash}}\ c{\isaliteral{2E}{\isachardot}}t\isaliteral{5C3C5E697375623E}{}\isactrlisub {\isadigit{1}}} with the side-conditions
  \isa{c\ {\isaliteral{5C3C6E6F7465713E}{\isasymnoteq}}\ d} and \isa{a\ {\isaliteral{5C3C6E6F7465713E}{\isasymnoteq}}\ d}. Using these side-condition, we can reduce our goal to establishing
  \isa{{\isaliteral{5C3C6E61626C613E}{\isasymnabla}}\ {\isaliteral{5C3C7475726E7374696C653E}{\isasymturnstile}}\ a\ {\isaliteral{23}{\isacharhash}}\ t\isaliteral{5C3C5E697375623E}{}\isactrlisub {\isadigit{2}}}. We can also discharge the case where \isa{a\ {\isaliteral{3D}{\isacharequal}}\ c}, since we know
  that \isa{{\isaliteral{5C3C6E61626C613E}{\isasymnabla}}\ {\isaliteral{5C3C7475726E7374696C653E}{\isasymturnstile}}\ c\ {\isaliteral{23}{\isacharhash}}\ t\isaliteral{5C3C5E697375623E}{}\isactrlisub {\isadigit{2}}} holds by the side-condition of (\isa{{\isaliteral{5C3C617070726F783E}{\isasymapprox}}}-abstraction$_2$).
  In case \isa{a\ {\isaliteral{5C3C6E6F7465713E}{\isasymnoteq}}\ c}, we can infer \isa{{\isaliteral{5C3C6E61626C613E}{\isasymnabla}}\ {\isaliteral{5C3C7475726E7374696C653E}{\isasymturnstile}}\ a\ {\isaliteral{23}{\isacharhash}}\ t\isaliteral{5C3C5E697375623E}{}\isactrlisub {\isadigit{1}}} from \isa{{\isaliteral{28}{\isacharparenleft}}{\isaliteral{2A}{\isacharasterisk}}{\isaliteral{29}{\isacharparenright}}}, and use 
  the induction hypothesis
  to conclude with \isa{{\isaliteral{5C3C6E61626C613E}{\isasymnabla}}\ {\isaliteral{5C3C7475726E7374696C653E}{\isasymturnstile}}\ a\ {\isaliteral{23}{\isacharhash}}\ {\isaliteral{28}{\isacharparenleft}}c\ d{\isaliteral{29}{\isacharparenright}}\ {\isaliteral{5C3C62756C6C65743E}{\isasymbullet}}\ t\isaliteral{5C3C5E697375623E}{}\isactrlisub {\isadigit{2}}}. Using Lemma~\ref{fresheqvt} we can infer
  that \isa{{\isaliteral{5C3C6E61626C613E}{\isasymnabla}}\ {\isaliteral{5C3C7475726E7374696C653E}{\isasymturnstile}}\ {\isaliteral{28}{\isacharparenleft}}c\ d{\isaliteral{29}{\isacharparenright}}\ {\isaliteral{5C3C62756C6C65743E}{\isasymbullet}}\ a\ {\isaliteral{23}{\isacharhash}}\ {\isaliteral{28}{\isacharparenleft}}c\ d{\isaliteral{29}{\isacharparenright}}{\isaliteral{28}{\isacharparenleft}}c\ d{\isaliteral{29}{\isacharparenright}}\ {\isaliteral{5C3C62756C6C65743E}{\isasymbullet}}\ t\isaliteral{5C3C5E697375623E}{}\isactrlisub {\isadigit{2}}} holds, whose left-hand side simplifies
  to just \isa{a} (we have that \isa{a\ {\isaliteral{5C3C6E6F7465713E}{\isasymnoteq}}\ d} and \isa{a\ {\isaliteral{5C3C6E6F7465713E}{\isasymnoteq}}\ c}). 
  For the right-hand side we can prove \isa{{\isaliteral{28}{\isacharparenleft}}c\ d{\isaliteral{29}{\isacharparenright}}{\isaliteral{28}{\isacharparenleft}}c\ d{\isaliteral{29}{\isacharparenright}}\ {\isaliteral{5C3C62756C6C65743E}{\isasymbullet}}\ t\isaliteral{5C3C5E697375623E}{}\isactrlisub {\isadigit{2}}\ {\isaliteral{5C3C73696D3E}{\isasymsim}}\ t\isaliteral{5C3C5E697375623E}{}\isactrlisub {\isadigit{2}}},
  since \isa{ds\ {\isaliteral{28}{\isacharparenleft}}{\isaliteral{28}{\isacharparenleft}}c\ d{\isaliteral{29}{\isacharparenright}}{\isaliteral{28}{\isacharparenleft}}c\ d{\isaliteral{29}{\isacharparenright}}{\isaliteral{29}{\isacharparenright}}\ {\isaliteral{5B}{\isacharbrackleft}}{\isaliteral{5D}{\isacharbrackright}}\ {\isaliteral{3D}{\isacharequal}}\ {\isaliteral{5C3C656D7074797365743E}{\isasymemptyset}}}. From this we can conclude this 
  case using Lemma~\ref{weak}\isa{{\isaliteral{28}{\isacharparenleft}}i{\isaliteral{29}{\isacharparenright}}}.
  \end{proof}

  \noindent
  The point in this proof is that without the weak equivalence and 
  without Lemma~\ref{weak}, we would need to perform many more 
  ``reshuffles'' of swappings than the single reference to \isa{{\isaliteral{5C3C73696D3E}{\isasymsim}}} in the 
  proof above \cite{UrbanPittsGabbay04}.
  The next property on the way to establish transitivity  proves
  the equivariance for \isa{{\isaliteral{5C3C617070726F783E}{\isasymapprox}}}.

  \begin{lemma}\label{equpi}
  \isa{{\normalsize{}If\,}\ \ {\isaliteral{5C3C6E61626C613E}{\isasymnabla}}\ {\isaliteral{5C3C7475726E7374696C653E}{\isasymturnstile}}\ t\isaliteral{5C3C5E697375623E}{}\isactrlisub {\isadigit{1}}\ {\isaliteral{5C3C617070726F783E}{\isasymapprox}}\ t\isaliteral{5C3C5E697375623E}{}\isactrlisub {\isadigit{2}}\ {\normalsize \,then\,}\ \ {\isaliteral{5C3C6E61626C613E}{\isasymnabla}}\ {\isaliteral{5C3C7475726E7374696C653E}{\isasymturnstile}}\ {\isaliteral{5C3C70693E}{\isasympi}}\ {\isaliteral{5C3C62756C6C65743E}{\isasymbullet}}\ t\isaliteral{5C3C5E697375623E}{}\isactrlisub {\isadigit{1}}\ {\isaliteral{5C3C617070726F783E}{\isasymapprox}}\ {\isaliteral{5C3C70693E}{\isasympi}}\ {\isaliteral{5C3C62756C6C65743E}{\isasymbullet}}\ t\isaliteral{5C3C5E697375623E}{}\isactrlisub {\isadigit{2}}{\isaliteral{2E}{\isachardot}}}
  \end{lemma}

  \noindent
  Also with this lemma the induction on \isa{{\isaliteral{5C3C617070726F783E}{\isasymapprox}}} does not go 
  through without the help of weak equivalence, because in the 
  (\isa{{\isaliteral{5C3C617070726F783E}{\isasymapprox}}}-abstraction$_2$)-case we need to show that \isa{{\isaliteral{5C3C6E61626C613E}{\isasymnabla}}\ {\isaliteral{5C3C7475726E7374696C653E}{\isasymturnstile}}\ {\isaliteral{5C3C70693E}{\isasympi}}\ {\isaliteral{5C3C62756C6C65743E}{\isasymbullet}}\ t\isaliteral{5C3C5E697375623E}{}\isactrlisub {\isadigit{1}}\ {\isaliteral{5C3C617070726F783E}{\isasymapprox}}\ {\isaliteral{5C3C70693E}{\isasympi}}\ {\isaliteral{5C3C62756C6C65743E}{\isasymbullet}}\ {\isaliteral{28}{\isacharparenleft}}a\ b{\isaliteral{29}{\isacharparenright}}\ {\isaliteral{5C3C62756C6C65743E}{\isasymbullet}}\ t\isaliteral{5C3C5E697375623E}{}\isactrlisub {\isadigit{2}}} implies
  \isa{{\isaliteral{5C3C6E61626C613E}{\isasymnabla}}\ {\isaliteral{5C3C7475726E7374696C653E}{\isasymturnstile}}\ {\isaliteral{5C3C70693E}{\isasympi}}\ {\isaliteral{5C3C62756C6C65743E}{\isasymbullet}}\ t\isaliteral{5C3C5E697375623E}{}\isactrlisub {\isadigit{1}}\ {\isaliteral{5C3C617070726F783E}{\isasymapprox}}\ {\isaliteral{28}{\isacharparenleft}}{\isaliteral{5C3C70693E}{\isasympi}}{\isaliteral{5C3C63646F743E}{\isasymcdot}}a\ \ {\isaliteral{5C3C70693E}{\isasympi}}{\isaliteral{5C3C63646F743E}{\isasymcdot}}b{\isaliteral{29}{\isacharparenright}}\ {\isaliteral{5C3C62756C6C65743E}{\isasymbullet}}\ {\isaliteral{5C3C70693E}{\isasympi}}\ {\isaliteral{5C3C62756C6C65743E}{\isasymbullet}}\ t\isaliteral{5C3C5E697375623E}{}\isactrlisub {\isadigit{2}}}. While it is easy to show that the right-hand sides are 
  equal, one cannot make use of this fact without a notion of transitivity. 

  \begin{proof}
  By induction on \isa{{\isaliteral{5C3C617070726F783E}{\isasymapprox}}}. The non-trivial case is the rule (\isa{{\isaliteral{5C3C617070726F783E}{\isasymapprox}}}-abstraction$_2$) where we know
  \isa{{\isaliteral{5C3C6E61626C613E}{\isasymnabla}}\ {\isaliteral{5C3C7475726E7374696C653E}{\isasymturnstile}}\ {\isaliteral{5C3C70693E}{\isasympi}}\ {\isaliteral{5C3C62756C6C65743E}{\isasymbullet}}\ t\isaliteral{5C3C5E697375623E}{}\isactrlisub {\isadigit{1}}\ {\isaliteral{5C3C617070726F783E}{\isasymapprox}}\ {\isaliteral{5C3C70693E}{\isasympi}}\ {\isaliteral{5C3C62756C6C65743E}{\isasymbullet}}\ {\isaliteral{28}{\isacharparenleft}}a\ b{\isaliteral{29}{\isacharparenright}}\ {\isaliteral{5C3C62756C6C65743E}{\isasymbullet}}\ t\isaliteral{5C3C5E697375623E}{}\isactrlisub {\isadigit{2}}} by induction hypothesis. We can show that
  \isa{{\isaliteral{5C3C70693E}{\isasympi}}\ {\isaliteral{40}{\isacharat}}\ {\isaliteral{28}{\isacharparenleft}}a\ b{\isaliteral{29}{\isacharparenright}}\ {\isaliteral{5C3C62756C6C65743E}{\isasymbullet}}\ t\isaliteral{5C3C5E697375623E}{}\isactrlisub {\isadigit{2}}\ {\isaliteral{5C3C73696D3E}{\isasymsim}}\ {\isaliteral{28}{\isacharparenleft}}{\isaliteral{5C3C70693E}{\isasympi}}{\isaliteral{5C3C63646F743E}{\isasymcdot}}a\ \ {\isaliteral{5C3C70693E}{\isasympi}}{\isaliteral{5C3C63646F743E}{\isasymcdot}}b{\isaliteral{29}{\isacharparenright}}\ {\isaliteral{40}{\isacharat}}\ {\isaliteral{5C3C70693E}{\isasympi}}\ {\isaliteral{5C3C62756C6C65743E}{\isasymbullet}}\ t\isaliteral{5C3C5E697375623E}{}\isactrlisub {\isadigit{2}}} holds (the corresponding disagreement set is empty). 
  Using Lemma~\ref{weak}\isa{{\isaliteral{28}{\isacharparenleft}}ii{\isaliteral{29}{\isacharparenright}}}, we can join both judgements and conclude with
  \isa{{\isaliteral{5C3C6E61626C613E}{\isasymnabla}}\ {\isaliteral{5C3C7475726E7374696C653E}{\isasymturnstile}}\ {\isaliteral{5C3C70693E}{\isasympi}}\ {\isaliteral{5C3C62756C6C65743E}{\isasymbullet}}\ t\isaliteral{5C3C5E697375623E}{}\isactrlisub {\isadigit{1}}\ {\isaliteral{5C3C617070726F783E}{\isasymapprox}}\ {\isaliteral{28}{\isacharparenleft}}{\isaliteral{5C3C70693E}{\isasympi}}{\isaliteral{5C3C63646F743E}{\isasymcdot}}a\ \ {\isaliteral{5C3C70693E}{\isasympi}}{\isaliteral{5C3C63646F743E}{\isasymcdot}}b{\isaliteral{29}{\isacharparenright}}\ {\isaliteral{5C3C62756C6C65743E}{\isasymbullet}}\ {\isaliteral{5C3C70693E}{\isasympi}}\ {\isaliteral{5C3C62756C6C65743E}{\isasymbullet}}\ t\isaliteral{5C3C5E697375623E}{}\isactrlisub {\isadigit{2}}}.
  \end{proof}

  \noindent
  The next lemma relates the freshness and equivalence relations.

  \begin{lemma}\label{dsequ}
  If \isa{{\isaliteral{5C3C666F72616C6C3E}{\isasymforall}}a{\isaliteral{5C3C696E3E}{\isasymin}}ds\ {\isaliteral{5C3C70693E}{\isasympi}}\ {\isaliteral{5C3C70693E}{\isasympi}}{\isaliteral{27}{\isacharprime}}{\isaliteral{2E}{\isachardot}}\ {\isaliteral{5C3C6E61626C613E}{\isasymnabla}}\ {\isaliteral{5C3C7475726E7374696C653E}{\isasymturnstile}}\ a\ {\isaliteral{23}{\isacharhash}}\ t} then
   \isa{\ {\isaliteral{5C3C6E61626C613E}{\isasymnabla}}\ {\isaliteral{5C3C7475726E7374696C653E}{\isasymturnstile}}\ {\isaliteral{5C3C70693E}{\isasympi}}\ {\isaliteral{5C3C62756C6C65743E}{\isasymbullet}}\ t\ {\isaliteral{5C3C617070726F783E}{\isasymapprox}}\ {\isaliteral{5C3C70693E}{\isasympi}}{\isaliteral{27}{\isacharprime}}\ {\isaliteral{5C3C62756C6C65743E}{\isasymbullet}}\ t}, and vice versa.
  \end{lemma}

  \begin{proof}
  By induction on \isa{t} generalising over the permutation \isa{{\isaliteral{5C3C70693E}{\isasympi}}{\isaliteral{27}{\isacharprime}}}. The generalisation 
  is needed in order to get the abstraction case through.
  \end{proof}

  \noindent
  The crucial lemma in \cite{KumarNorrish10}, which will allow us to prove the 
  transitivity property by a straightforward rule induction, is the next one. Its proof still
  needs to analyse several cases, but the reasoning is much simpler than in the proof by 
  induction over the size of terms in \cite{UrbanPittsGabbay04}. 
 
  \begin{lemma}\label{equtranspi}
  \isa{{\normalsize{}If\,}\ \mbox{\ {\isaliteral{5C3C6E61626C613E}{\isasymnabla}}\ {\isaliteral{5C3C7475726E7374696C653E}{\isasymturnstile}}\ t\isaliteral{5C3C5E697375623E}{}\isactrlisub {\isadigit{1}}\ {\isaliteral{5C3C617070726F783E}{\isasymapprox}}\ t\isaliteral{5C3C5E697375623E}{}\isactrlisub {\isadigit{2}}}\ {\normalsize \,and\,}\ \mbox{\ {\isaliteral{5C3C6E61626C613E}{\isasymnabla}}\ {\isaliteral{5C3C7475726E7374696C653E}{\isasymturnstile}}\ t\isaliteral{5C3C5E697375623E}{}\isactrlisub {\isadigit{2}}\ {\isaliteral{5C3C617070726F783E}{\isasymapprox}}\ {\isaliteral{5C3C70693E}{\isasympi}}\ {\isaliteral{5C3C62756C6C65743E}{\isasymbullet}}\ t\isaliteral{5C3C5E697375623E}{}\isactrlisub {\isadigit{2}}}\ {\normalsize \,then\,}\ \ {\isaliteral{5C3C6E61626C613E}{\isasymnabla}}\ {\isaliteral{5C3C7475726E7374696C653E}{\isasymturnstile}}\ t\isaliteral{5C3C5E697375623E}{}\isactrlisub {\isadigit{1}}\ {\isaliteral{5C3C617070726F783E}{\isasymapprox}}\ {\isaliteral{5C3C70693E}{\isasympi}}\ {\isaliteral{5C3C62756C6C65743E}{\isasymbullet}}\ t\isaliteral{5C3C5E697375623E}{}\isactrlisub {\isadigit{2}}{\isaliteral{2E}{\isachardot}}} 
  \end{lemma}
  
  \begin{proof}
  By induction on the first \isa{{\isaliteral{5C3C617070726F783E}{\isasymapprox}}}-judgement with a generalisation over \isa{{\isaliteral{5C3C70693E}{\isasympi}}}. The interesting 
  case is (\isa{{\isaliteral{5C3C617070726F783E}{\isasymapprox}}}-abstraction$_2$). We know \isa{{\isaliteral{5C3C6E61626C613E}{\isasymnabla}}\ {\isaliteral{5C3C7475726E7374696C653E}{\isasymturnstile}}\ b{\isaliteral{2E}{\isachardot}}t\isaliteral{5C3C5E697375623E}{}\isactrlisub {\isadigit{2}}\ {\isaliteral{5C3C617070726F783E}{\isasymapprox}}\ {\isaliteral{28}{\isacharparenleft}}{\isaliteral{5C3C70693E}{\isasympi}}\ {\isaliteral{5C3C62756C6C65743E}{\isasymbullet}}\ b{\isaliteral{29}{\isacharparenright}}{\isaliteral{2E}{\isachardot}}{\isaliteral{28}{\isacharparenleft}}{\isaliteral{5C3C70693E}{\isasympi}}\ {\isaliteral{5C3C62756C6C65743E}{\isasymbullet}}\ t\isaliteral{5C3C5E697375623E}{}\isactrlisub {\isadigit{2}}{\isaliteral{29}{\isacharparenright}}} and have
  to prove \isa{{\isaliteral{5C3C6E61626C613E}{\isasymnabla}}\ {\isaliteral{5C3C7475726E7374696C653E}{\isasymturnstile}}\ a{\isaliteral{2E}{\isachardot}}t\isaliteral{5C3C5E697375623E}{}\isactrlisub {\isadigit{1}}\ {\isaliteral{5C3C617070726F783E}{\isasymapprox}}\ {\isaliteral{28}{\isacharparenleft}}{\isaliteral{5C3C70693E}{\isasympi}}\ {\isaliteral{5C3C62756C6C65743E}{\isasymbullet}}\ b{\isaliteral{29}{\isacharparenright}}{\isaliteral{2E}{\isachardot}}{\isaliteral{28}{\isacharparenleft}}{\isaliteral{5C3C70693E}{\isasympi}}\ {\isaliteral{5C3C62756C6C65743E}{\isasymbullet}}\ t\isaliteral{5C3C5E697375623E}{}\isactrlisub {\isadigit{2}}{\isaliteral{29}{\isacharparenright}}} with \isa{a\ {\isaliteral{5C3C6E6F7465713E}{\isasymnoteq}}\ b}. We have to analyse
  several cases about \isa{a} equal equal with \isa{{\isaliteral{5C3C70693E}{\isasympi}}\ {\isaliteral{5C3C62756C6C65743E}{\isasymbullet}}\ b}, and \isa{b} being equal 
  with \isa{{\isaliteral{5C3C70693E}{\isasympi}}\ {\isaliteral{5C3C62756C6C65743E}{\isasymbullet}}\ b}. Let us give the details for the case \isa{a\ {\isaliteral{5C3C6E6F7465713E}{\isasymnoteq}}\ {\isaliteral{5C3C70693E}{\isasympi}}\ {\isaliteral{5C3C62756C6C65743E}{\isasymbullet}}\ b} and \isa{b\ {\isaliteral{5C3C6E6F7465713E}{\isasymnoteq}}\ {\isaliteral{5C3C70693E}{\isasympi}}\ {\isaliteral{5C3C62756C6C65743E}{\isasymbullet}}\ b}.
  From the assumption we can infer \isa{{\isaliteral{28}{\isacharparenleft}}{\isaliteral{2A}{\isacharasterisk}}{\isaliteral{29}{\isacharparenright}}} \isa{{\isaliteral{5C3C6E61626C613E}{\isasymnabla}}\ {\isaliteral{5C3C7475726E7374696C653E}{\isasymturnstile}}\ b\ {\isaliteral{23}{\isacharhash}}\ {\isaliteral{5C3C70693E}{\isasympi}}\ {\isaliteral{5C3C62756C6C65743E}{\isasymbullet}}\ t\isaliteral{5C3C5E697375623E}{}\isactrlisub {\isadigit{2}}} and 
  \isa{{\isaliteral{28}{\isacharparenleft}}{\isaliteral{2A}{\isacharasterisk}}{\isaliteral{2A}{\isacharasterisk}}{\isaliteral{29}{\isacharparenright}}} \isa{{\isaliteral{5C3C6E61626C613E}{\isasymnabla}}\ {\isaliteral{5C3C7475726E7374696C653E}{\isasymturnstile}}\ t\isaliteral{5C3C5E697375623E}{}\isactrlisub {\isadigit{2}}\ {\isaliteral{5C3C617070726F783E}{\isasymapprox}}\ {\isaliteral{28}{\isacharparenleft}}b\ {\isaliteral{5C3C70693E}{\isasympi}}{\isaliteral{5C3C63646F743E}{\isasymcdot}}b{\isaliteral{29}{\isacharparenright}}\ {\isaliteral{5C3C62756C6C65743E}{\isasymbullet}}\ {\isaliteral{5C3C70693E}{\isasympi}}\ {\isaliteral{5C3C62756C6C65743E}{\isasymbullet}}\ t\isaliteral{5C3C5E697375623E}{}\isactrlisub {\isadigit{2}}}.
  The side-condition on the first judgement gives us \isa{{\isaliteral{5C3C6E61626C613E}{\isasymnabla}}\ {\isaliteral{5C3C7475726E7374696C653E}{\isasymturnstile}}\ a\ {\isaliteral{23}{\isacharhash}}\ t\isaliteral{5C3C5E697375623E}{}\isactrlisub {\isadigit{2}}}. 
  We have to show \isa{{\isaliteral{5C3C6E61626C613E}{\isasymnabla}}\ {\isaliteral{5C3C7475726E7374696C653E}{\isasymturnstile}}\ a\ {\isaliteral{23}{\isacharhash}}\ {\isaliteral{5C3C70693E}{\isasympi}}\ {\isaliteral{5C3C62756C6C65743E}{\isasymbullet}}\ t\isaliteral{5C3C5E697375623E}{}\isactrlisub {\isadigit{2}}} and \isa{{\isaliteral{5C3C6E61626C613E}{\isasymnabla}}\ {\isaliteral{5C3C7475726E7374696C653E}{\isasymturnstile}}\ t\isaliteral{5C3C5E697375623E}{}\isactrlisub {\isadigit{1}}\ {\isaliteral{5C3C617070726F783E}{\isasymapprox}}\ {\isaliteral{28}{\isacharparenleft}}a\ {\isaliteral{5C3C70693E}{\isasympi}}{\isaliteral{5C3C63646F743E}{\isasymcdot}}b{\isaliteral{29}{\isacharparenright}}\ {\isaliteral{5C3C62756C6C65743E}{\isasymbullet}}\ {\isaliteral{5C3C70693E}{\isasympi}}\ {\isaliteral{5C3C62756C6C65743E}{\isasymbullet}}\ t\isaliteral{5C3C5E697375623E}{}\isactrlisub {\isadigit{2}}}. 
  To infer the first fact, we use \isa{{\isaliteral{5C3C6E61626C613E}{\isasymnabla}}\ {\isaliteral{5C3C7475726E7374696C653E}{\isasymturnstile}}\ a\ {\isaliteral{23}{\isacharhash}}\ t\isaliteral{5C3C5E697375623E}{}\isactrlisub {\isadigit{2}}} together with \isa{{\isaliteral{28}{\isacharparenleft}}{\isaliteral{2A}{\isacharasterisk}}{\isaliteral{2A}{\isacharasterisk}}{\isaliteral{29}{\isacharparenright}}} and 
  Lemmas~\ref{freshequ} and \ref{fresheqvt}. For the second, the
  induction hypothesis states that for any \isa{{\isaliteral{5C3C70693E}{\isasympi}}} we have \isa{{\isaliteral{5C3C6E61626C613E}{\isasymnabla}}\ {\isaliteral{5C3C7475726E7374696C653E}{\isasymturnstile}}\ t\isaliteral{5C3C5E697375623E}{}\isactrlisub {\isadigit{1}}\ {\isaliteral{5C3C617070726F783E}{\isasymapprox}}\ {\isaliteral{5C3C70693E}{\isasympi}}\ {\isaliteral{5C3C62756C6C65743E}{\isasymbullet}}\ {\isaliteral{28}{\isacharparenleft}}a\ b{\isaliteral{29}{\isacharparenright}}\ {\isaliteral{5C3C62756C6C65743E}{\isasymbullet}}\ t\isaliteral{5C3C5E697375623E}{}\isactrlisub {\isadigit{2}}} 
  provided
  \isa{{\isaliteral{5C3C6E61626C613E}{\isasymnabla}}\ {\isaliteral{5C3C7475726E7374696C653E}{\isasymturnstile}}\ {\isaliteral{28}{\isacharparenleft}}a\ b{\isaliteral{29}{\isacharparenright}}\ {\isaliteral{5C3C62756C6C65743E}{\isasymbullet}}\ t\isaliteral{5C3C5E697375623E}{}\isactrlisub {\isadigit{2}}\ {\isaliteral{5C3C617070726F783E}{\isasymapprox}}\ {\isaliteral{5C3C70693E}{\isasympi}}\ {\isaliteral{5C3C62756C6C65743E}{\isasymbullet}}\ {\isaliteral{28}{\isacharparenleft}}a\ b{\isaliteral{29}{\isacharparenright}}\ {\isaliteral{5C3C62756C6C65743E}{\isasymbullet}}\ t\isaliteral{5C3C5E697375623E}{}\isactrlisub {\isadigit{2}}} holds. We use the induction hypothesis with the 
  permutation \isa{{\isaliteral{5C3C70693E}{\isasympi}}\ {\isaliteral{5C3C65717569763E}{\isasymequiv}}\ {\isaliteral{28}{\isacharparenleft}}a\ {\isaliteral{5C3C70693E}{\isasympi}}{\isaliteral{5C3C63646F743E}{\isasymcdot}}b{\isaliteral{29}{\isacharparenright}}\ {\isaliteral{40}{\isacharat}}\ {\isaliteral{5C3C70693E}{\isasympi}}\ {\isaliteral{40}{\isacharat}}\ {\isaliteral{28}{\isacharparenleft}}a\ b{\isaliteral{29}{\isacharparenright}}}. This means after simplification the
  precondition of the IH we need to establish is \isa{{\isaliteral{28}{\isacharparenleft}}{\isaliteral{2A}{\isacharasterisk}}{\isaliteral{2A}{\isacharasterisk}}{\isaliteral{2A}{\isacharasterisk}}{\isaliteral{29}{\isacharparenright}}} 
   \isa{{\isaliteral{5C3C6E61626C613E}{\isasymnabla}}\ {\isaliteral{5C3C7475726E7374696C653E}{\isasymturnstile}}\ {\isaliteral{28}{\isacharparenleft}}a\ b{\isaliteral{29}{\isacharparenright}}\ {\isaliteral{5C3C62756C6C65743E}{\isasymbullet}}\ t\isaliteral{5C3C5E697375623E}{}\isactrlisub {\isadigit{2}}\ {\isaliteral{5C3C617070726F783E}{\isasymapprox}}\ \ {\isaliteral{28}{\isacharparenleft}}a\ {\isaliteral{5C3C70693E}{\isasympi}}{\isaliteral{5C3C63646F743E}{\isasymcdot}}b{\isaliteral{29}{\isacharparenright}}\ {\isaliteral{5C3C62756C6C65743E}{\isasymbullet}}\ {\isaliteral{5C3C70693E}{\isasympi}}\ {\isaliteral{5C3C62756C6C65743E}{\isasymbullet}}\ t\isaliteral{5C3C5E697375623E}{}\isactrlisub {\isadigit{2}}}.
  By Lemma~\ref{dsequ} we can transform
  \isa{{\isaliteral{28}{\isacharparenleft}}{\isaliteral{2A}{\isacharasterisk}}{\isaliteral{2A}{\isacharasterisk}}{\isaliteral{29}{\isacharparenright}}} to  \isa{{\isaliteral{5C3C666F72616C6C3E}{\isasymforall}}c\ {\isaliteral{5C3C696E3E}{\isasymin}}\ ds\ {\isaliteral{5B}{\isacharbrackleft}}{\isaliteral{5D}{\isacharbrackright}}\ {\isaliteral{28}{\isacharparenleft}}{\isaliteral{28}{\isacharparenleft}}b{\isaliteral{2C}{\isacharcomma}}\ {\isaliteral{5C3C70693E}{\isasympi}}{\isaliteral{5C3C63646F743E}{\isasymcdot}}b{\isaliteral{29}{\isacharparenright}}\ {\isaliteral{40}{\isacharat}}\ {\isaliteral{5C3C70693E}{\isasympi}}{\isaliteral{29}{\isacharparenright}}{\isaliteral{2E}{\isachardot}}\ \ {\isaliteral{5C3C6E61626C613E}{\isasymnabla}}\ {\isaliteral{5C3C7475726E7374696C653E}{\isasymturnstile}}\ c\ {\isaliteral{23}{\isacharhash}}\ t\isaliteral{5C3C5E697375623E}{}\isactrlisub {\isadigit{2}}}. Similarly with \isa{{\isaliteral{28}{\isacharparenleft}}{\isaliteral{2A}{\isacharasterisk}}{\isaliteral{2A}{\isacharasterisk}}{\isaliteral{2A}{\isacharasterisk}}{\isaliteral{29}{\isacharparenright}}}.
  Furthermore we can show that

  \[
  \isa{ds\ {\isaliteral{28}{\isacharparenleft}}a\ b{\isaliteral{29}{\isacharparenright}}\ {\isaliteral{28}{\isacharparenleft}}{\isaliteral{28}{\isacharparenleft}}a\ {\isaliteral{5C3C70693E}{\isasympi}}{\isaliteral{5C3C63646F743E}{\isasymcdot}}b{\isaliteral{29}{\isacharparenright}}\ {\isaliteral{40}{\isacharat}}\ {\isaliteral{5C3C70693E}{\isasympi}}{\isaliteral{29}{\isacharparenright}}\ {\isaliteral{5C3C73756273657465713E}{\isasymsubseteq}}\ ds\ {\isaliteral{5B}{\isacharbrackleft}}{\isaliteral{5D}{\isacharbrackright}}\ {\isaliteral{28}{\isacharparenleft}}{\isaliteral{28}{\isacharparenleft}}b\ {\isaliteral{5C3C70693E}{\isasympi}}{\isaliteral{5C3C63646F743E}{\isasymcdot}}b{\isaliteral{29}{\isacharparenright}}\ {\isaliteral{40}{\isacharat}}\ {\isaliteral{5C3C70693E}{\isasympi}}{\isaliteral{29}{\isacharparenright}}\ {\isaliteral{5C3C756E696F6E3E}{\isasymunion}}\ {\isaliteral{7B}{\isacharbraceleft}}a{\isaliteral{2C}{\isacharcomma}}\ {\isaliteral{5C3C70693E}{\isasympi}}{\isaliteral{5C3C63646F743E}{\isasymcdot}}b{\isaliteral{7D}{\isacharbraceright}}}
  \]

  \noindent
  holds. This means it remains to show that \isa{{\isaliteral{5C3C6E61626C613E}{\isasymnabla}}\ {\isaliteral{5C3C7475726E7374696C653E}{\isasymturnstile}}\ a\ {\isaliteral{23}{\isacharhash}}\ t\isaliteral{5C3C5E697375623E}{}\isactrlisub {\isadigit{2}}} (which we already inferred above)
  and \isa{{\isaliteral{5C3C6E61626C613E}{\isasymnabla}}\ {\isaliteral{5C3C7475726E7374696C653E}{\isasymturnstile}}\ {\isaliteral{5C3C70693E}{\isasympi}}{\isaliteral{5C3C63646F743E}{\isasymcdot}}b\ {\isaliteral{23}{\isacharhash}}\ t\isaliteral{5C3C5E697375623E}{}\isactrlisub {\isadigit{2}}} hold. For the latter, we consider the cases \isa{b\ {\isaliteral{3D}{\isacharequal}}\ {\isaliteral{5C3C70693E}{\isasympi}}\ {\isaliteral{5C3C62756C6C65743E}{\isasymbullet}}\ {\isaliteral{5C3C70693E}{\isasympi}}\ {\isaliteral{5C3C62756C6C65743E}{\isasymbullet}}\ b} and
  \isa{b\ {\isaliteral{5C3C6E6F7465713E}{\isasymnoteq}}\ {\isaliteral{5C3C70693E}{\isasympi}}\ {\isaliteral{5C3C62756C6C65743E}{\isasymbullet}}\ {\isaliteral{5C3C70693E}{\isasympi}}\ {\isaliteral{5C3C62756C6C65743E}{\isasymbullet}}\ b}. In the first case we infer \isa{{\isaliteral{5C3C6E61626C613E}{\isasymnabla}}\ {\isaliteral{5C3C7475726E7374696C653E}{\isasymturnstile}}\ {\isaliteral{5C3C70693E}{\isasympi}}{\isaliteral{5C3C63646F743E}{\isasymcdot}}b\ {\isaliteral{23}{\isacharhash}}\ t\isaliteral{5C3C5E697375623E}{}\isactrlisub {\isadigit{2}}} from \isa{{\isaliteral{28}{\isacharparenleft}}{\isaliteral{2A}{\isacharasterisk}}{\isaliteral{29}{\isacharparenright}}}
  using Lemma~\ref{fresheqvt}. In the second case we have that \isa{{\isaliteral{5C3C70693E}{\isasympi}}\ {\isaliteral{5C3C62756C6C65743E}{\isasymbullet}}\ b\ {\isaliteral{5C3C696E3E}{\isasymin}}\ ds\ {\isaliteral{5B}{\isacharbrackleft}}{\isaliteral{5D}{\isacharbrackright}}\ {\isaliteral{28}{\isacharparenleft}}{\isaliteral{28}{\isacharparenleft}}b\ {\isaliteral{5C3C70693E}{\isasympi}}{\isaliteral{5C3C63646F743E}{\isasymcdot}}b{\isaliteral{29}{\isacharparenright}}\ {\isaliteral{40}{\isacharat}}\ {\isaliteral{5C3C70693E}{\isasympi}}{\isaliteral{29}{\isacharparenright}}}.
  So finally we can use the induction hypothesis, which simplified gives us 
  \isa{{\isaliteral{5C3C6E61626C613E}{\isasymnabla}}\ {\isaliteral{5C3C7475726E7374696C653E}{\isasymturnstile}}\ t\isaliteral{5C3C5E697375623E}{}\isactrlisub {\isadigit{1}}\ {\isaliteral{5C3C617070726F783E}{\isasymapprox}}\ {\isaliteral{28}{\isacharparenleft}}a\ {\isaliteral{5C3C70693E}{\isasympi}}{\isaliteral{5C3C63646F743E}{\isasymcdot}}b{\isaliteral{29}{\isacharparenright}}\ {\isaliteral{5C3C62756C6C65743E}{\isasymbullet}}\ {\isaliteral{5C3C70693E}{\isasympi}}\ {\isaliteral{5C3C62756C6C65743E}{\isasymbullet}}\ t\isaliteral{5C3C5E697375623E}{}\isactrlisub {\isadigit{2}}} as needed.
  \end{proof}

  \noindent
  With this lemma under our belt, we are finally in the position to prove the transitivity property.

  \begin{lemma}\label{equtrans}
  \isa{{\normalsize{}If\,}\ \mbox{\ {\isaliteral{5C3C6E61626C613E}{\isasymnabla}}\ {\isaliteral{5C3C7475726E7374696C653E}{\isasymturnstile}}\ t\isaliteral{5C3C5E697375623E}{}\isactrlisub {\isadigit{1}}\ {\isaliteral{5C3C617070726F783E}{\isasymapprox}}\ t\isaliteral{5C3C5E697375623E}{}\isactrlisub {\isadigit{2}}}\ {\normalsize \,and\,}\ \mbox{\ {\isaliteral{5C3C6E61626C613E}{\isasymnabla}}\ {\isaliteral{5C3C7475726E7374696C653E}{\isasymturnstile}}\ t\isaliteral{5C3C5E697375623E}{}\isactrlisub {\isadigit{2}}\ {\isaliteral{5C3C617070726F783E}{\isasymapprox}}\ t\isaliteral{5C3C5E697375623E}{}\isactrlisub {\isadigit{3}}}\ {\normalsize \,then\,}\ \ {\isaliteral{5C3C6E61626C613E}{\isasymnabla}}\ {\isaliteral{5C3C7475726E7374696C653E}{\isasymturnstile}}\ t\isaliteral{5C3C5E697375623E}{}\isactrlisub {\isadigit{1}}\ {\isaliteral{5C3C617070726F783E}{\isasymapprox}}\ t\isaliteral{5C3C5E697375623E}{}\isactrlisub {\isadigit{3}}{\isaliteral{2E}{\isachardot}}}
  \end{lemma}
  
  \begin{proof}
  By induction on the first judgement generalising over \isa{t\isaliteral{5C3C5E697375623E}{}\isactrlisub {\isadigit{3}}}. We then analyse the possible
  instances for the second judgement. The non-trivial case is where both judgements are instances
  of the rule (\isa{{\isaliteral{5C3C617070726F783E}{\isasymapprox}}}-abstraction$_2$). We have \isa{{\isaliteral{5C3C6E61626C613E}{\isasymnabla}}\ {\isaliteral{5C3C7475726E7374696C653E}{\isasymturnstile}}\ t\isaliteral{5C3C5E697375623E}{}\isactrlisub {\isadigit{1}}\ {\isaliteral{5C3C617070726F783E}{\isasymapprox}}\ {\isaliteral{28}{\isacharparenleft}}a\ b{\isaliteral{29}{\isacharparenright}}\ {\isaliteral{5C3C62756C6C65743E}{\isasymbullet}}\ t\isaliteral{5C3C5E697375623E}{}\isactrlisub {\isadigit{2}}} and
  \isa{{\isaliteral{28}{\isacharparenleft}}{\isaliteral{2A}{\isacharasterisk}}{\isaliteral{29}{\isacharparenright}}} \isa{{\isaliteral{5C3C6E61626C613E}{\isasymnabla}}\ {\isaliteral{5C3C7475726E7374696C653E}{\isasymturnstile}}\ t\isaliteral{5C3C5E697375623E}{}\isactrlisub {\isadigit{2}}\ {\isaliteral{5C3C617070726F783E}{\isasymapprox}}\ {\isaliteral{28}{\isacharparenleft}}b\ c{\isaliteral{29}{\isacharparenright}}\ {\isaliteral{5C3C62756C6C65743E}{\isasymbullet}}\ t\isaliteral{5C3C5E697375623E}{}\isactrlisub {\isadigit{3}}} with \isa{a}, \isa{b} and \isa{c} being distinct. 
  We also have \isa{{\isaliteral{28}{\isacharparenleft}}{\isaliteral{2A}{\isacharasterisk}}{\isaliteral{2A}{\isacharasterisk}}{\isaliteral{29}{\isacharparenright}}} \isa{{\isaliteral{5C3C6E61626C613E}{\isasymnabla}}\ {\isaliteral{5C3C7475726E7374696C653E}{\isasymturnstile}}\ a\ {\isaliteral{23}{\isacharhash}}\ t\isaliteral{5C3C5E697375623E}{}\isactrlisub {\isadigit{2}}} and \isa{{\isaliteral{28}{\isacharparenleft}}{\isaliteral{2A}{\isacharasterisk}}{\isaliteral{2A}{\isacharasterisk}}{\isaliteral{2A}{\isacharasterisk}}{\isaliteral{29}{\isacharparenright}}} \isa{{\isaliteral{5C3C6E61626C613E}{\isasymnabla}}\ {\isaliteral{5C3C7475726E7374696C653E}{\isasymturnstile}}\ b\ {\isaliteral{23}{\isacharhash}}\ t\isaliteral{5C3C5E697375623E}{}\isactrlisub {\isadigit{3}}}. 
  We have to show \isa{{\isaliteral{5C3C6E61626C613E}{\isasymnabla}}\ {\isaliteral{5C3C7475726E7374696C653E}{\isasymturnstile}}\ a\ {\isaliteral{23}{\isacharhash}}\ t\isaliteral{5C3C5E697375623E}{}\isactrlisub {\isadigit{3}}} and 
  \isa{{\isaliteral{5C3C6E61626C613E}{\isasymnabla}}\ {\isaliteral{5C3C7475726E7374696C653E}{\isasymturnstile}}\ t\isaliteral{5C3C5E697375623E}{}\isactrlisub {\isadigit{1}}\ {\isaliteral{5C3C617070726F783E}{\isasymapprox}}\ {\isaliteral{28}{\isacharparenleft}}a\ c{\isaliteral{29}{\isacharparenright}}\ {\isaliteral{5C3C62756C6C65743E}{\isasymbullet}}\ t\isaliteral{5C3C5E697375623E}{}\isactrlisub {\isadigit{3}}}. The first fact is a simple consequence of \isa{{\isaliteral{28}{\isacharparenleft}}{\isaliteral{2A}{\isacharasterisk}}{\isaliteral{29}{\isacharparenright}}} and 
  the Lemmas~\ref{fresheqvt} and 
  \ref{freshequ}. For the other case we can use the induction hypothesis to infer our
  proof obligation, provided we can establish that \isa{{\isaliteral{5C3C6E61626C613E}{\isasymnabla}}\ {\isaliteral{5C3C7475726E7374696C653E}{\isasymturnstile}}\ {\isaliteral{28}{\isacharparenleft}}a\ b{\isaliteral{29}{\isacharparenright}}\ {\isaliteral{5C3C62756C6C65743E}{\isasymbullet}}\ t\isaliteral{5C3C5E697375623E}{}\isactrlisub {\isadigit{2}}\ {\isaliteral{5C3C617070726F783E}{\isasymapprox}}\ {\isaliteral{28}{\isacharparenleft}}a\ c{\isaliteral{29}{\isacharparenright}}\ {\isaliteral{5C3C62756C6C65743E}{\isasymbullet}}\ t\isaliteral{5C3C5E697375623E}{}\isactrlisub {\isadigit{3}}} holds.
  From \isa{{\isaliteral{28}{\isacharparenleft}}{\isaliteral{2A}{\isacharasterisk}}{\isaliteral{29}{\isacharparenright}}} we have \isa{{\isaliteral{5C3C6E61626C613E}{\isasymnabla}}\ {\isaliteral{5C3C7475726E7374696C653E}{\isasymturnstile}}\ {\isaliteral{28}{\isacharparenleft}}a\ b{\isaliteral{29}{\isacharparenright}}\ {\isaliteral{5C3C62756C6C65743E}{\isasymbullet}}\ t\isaliteral{5C3C5E697375623E}{}\isactrlisub {\isadigit{2}}\ {\isaliteral{5C3C617070726F783E}{\isasymapprox}}\ {\isaliteral{28}{\isacharparenleft}}a\ b{\isaliteral{29}{\isacharparenright}}{\isaliteral{28}{\isacharparenleft}}b\ c{\isaliteral{29}{\isacharparenright}}\ {\isaliteral{5C3C62756C6C65743E}{\isasymbullet}}\ t\isaliteral{5C3C5E697375623E}{}\isactrlisub {\isadigit{3}}} using Lemma~\ref{equpi}.
  We also establish that \isa{{\isaliteral{5C3C6E61626C613E}{\isasymnabla}}\ {\isaliteral{5C3C7475726E7374696C653E}{\isasymturnstile}}\ {\isaliteral{28}{\isacharparenleft}}a\ b{\isaliteral{29}{\isacharparenright}}{\isaliteral{28}{\isacharparenleft}}b\ c{\isaliteral{29}{\isacharparenright}}\ {\isaliteral{5C3C62756C6C65743E}{\isasymbullet}}\ t\isaliteral{5C3C5E697375623E}{}\isactrlisub {\isadigit{3}}\ {\isaliteral{5C3C617070726F783E}{\isasymapprox}}\ {\isaliteral{28}{\isacharparenleft}}b\ c{\isaliteral{29}{\isacharparenright}}{\isaliteral{28}{\isacharparenleft}}a\ b{\isaliteral{29}{\isacharparenright}}{\isaliteral{28}{\isacharparenleft}}b\ c{\isaliteral{29}{\isacharparenright}}\ {\isaliteral{5C3C62756C6C65743E}{\isasymbullet}}\ t\isaliteral{5C3C5E697375623E}{}\isactrlisub {\isadigit{3}}} holds. 
  By Lemma~\ref{dsequ} we have to show that all atoms in the disagreement set are fresh
  w.r.t.~\isa{t\isaliteral{5C3C5E697375623E}{}\isactrlisub {\isadigit{3}}}. The disagreement set is equal to \isa{{\isaliteral{7B}{\isacharbraceleft}}a{\isaliteral{2C}{\isacharcomma}}\ b{\isaliteral{7D}{\isacharbraceright}}}. For \isa{b} the 
  property follows from \isa{{\isaliteral{28}{\isacharparenleft}}{\isaliteral{2A}{\isacharasterisk}}{\isaliteral{2A}{\isacharasterisk}}{\isaliteral{2A}{\isacharasterisk}}{\isaliteral{29}{\isacharparenright}}}.  For \isa{a} we use \isa{{\isaliteral{28}{\isacharparenleft}}{\isaliteral{2A}{\isacharasterisk}}{\isaliteral{29}{\isacharparenright}}} and \isa{{\isaliteral{28}{\isacharparenleft}}{\isaliteral{2A}{\isacharasterisk}}{\isaliteral{2A}{\isacharasterisk}}{\isaliteral{29}{\isacharparenright}}}. 
  So we can use Lemma~\ref{equtranspi} to infer \isa{{\isaliteral{28}{\isacharparenleft}}{\isaliteral{2A}{\isacharasterisk}}{\isaliteral{2A}{\isacharasterisk}}{\isaliteral{2A}{\isacharasterisk}}{\isaliteral{2A}{\isacharasterisk}}{\isaliteral{29}{\isacharparenright}}} 
  \isa{{\isaliteral{5C3C6E61626C613E}{\isasymnabla}}\ {\isaliteral{5C3C7475726E7374696C653E}{\isasymturnstile}}\ {\isaliteral{28}{\isacharparenleft}}a\ b{\isaliteral{29}{\isacharparenright}}\ {\isaliteral{5C3C62756C6C65743E}{\isasymbullet}}\ t\isaliteral{5C3C5E697375623E}{}\isactrlisub {\isadigit{2}}\ {\isaliteral{5C3C617070726F783E}{\isasymapprox}}\ {\isaliteral{28}{\isacharparenleft}}b\ c{\isaliteral{29}{\isacharparenright}}{\isaliteral{28}{\isacharparenleft}}a\ b{\isaliteral{29}{\isacharparenright}}{\isaliteral{28}{\isacharparenleft}}b\ c{\isaliteral{29}{\isacharparenright}}\ {\isaliteral{5C3C62756C6C65743E}{\isasymbullet}}\ t\isaliteral{5C3C5E697375623E}{}\isactrlisub {\isadigit{3}}}.
  It remains to show that \isa{{\isaliteral{5C3C6E61626C613E}{\isasymnabla}}\ {\isaliteral{5C3C7475726E7374696C653E}{\isasymturnstile}}\ {\isaliteral{28}{\isacharparenleft}}a\ b{\isaliteral{29}{\isacharparenright}}\ {\isaliteral{5C3C62756C6C65743E}{\isasymbullet}}\ t\isaliteral{5C3C5E697375623E}{}\isactrlisub {\isadigit{2}}\ {\isaliteral{5C3C617070726F783E}{\isasymapprox}}\ {\isaliteral{28}{\isacharparenleft}}a\ c{\isaliteral{29}{\isacharparenright}}\ {\isaliteral{5C3C62756C6C65743E}{\isasymbullet}}\ t\isaliteral{5C3C5E697375623E}{}\isactrlisub {\isadigit{3}}} holds. We can do so by
  using \isa{{\isaliteral{28}{\isacharparenleft}}{\isaliteral{2A}{\isacharasterisk}}{\isaliteral{2A}{\isacharasterisk}}{\isaliteral{2A}{\isacharasterisk}}{\isaliteral{2A}{\isacharasterisk}}{\isaliteral{29}{\isacharparenright}}} and Lemma~\ref{weak}, and showing 
  that \isa{{\isaliteral{28}{\isacharparenleft}}b\ c{\isaliteral{29}{\isacharparenright}}{\isaliteral{28}{\isacharparenleft}}a\ b{\isaliteral{29}{\isacharparenright}}{\isaliteral{28}{\isacharparenleft}}b\ c{\isaliteral{29}{\isacharparenright}}\ {\isaliteral{5C3C62756C6C65743E}{\isasymbullet}}\ t\isaliteral{5C3C5E697375623E}{}\isactrlisub {\isadigit{3}}\ {\isaliteral{5C3C73696D3E}{\isasymsim}}\ {\isaliteral{28}{\isacharparenleft}}a\ c{\isaliteral{29}{\isacharparenright}}\ {\isaliteral{5C3C62756C6C65743E}{\isasymbullet}}\ t\isaliteral{5C3C5E697375623E}{}\isactrlisub {\isadigit{3}}} holds. This in turn follows from the fact 
  that the disagreement set \isa{ds\ {\isaliteral{28}{\isacharparenleft}}{\isaliteral{28}{\isacharparenleft}}b\ c{\isaliteral{29}{\isacharparenright}}{\isaliteral{28}{\isacharparenleft}}a\ b{\isaliteral{29}{\isacharparenright}}{\isaliteral{28}{\isacharparenleft}}b\ c{\isaliteral{29}{\isacharparenright}}{\isaliteral{29}{\isacharparenright}}\ {\isaliteral{28}{\isacharparenleft}}a\ c{\isaliteral{29}{\isacharparenright}}} is empty. This concludes the case. 
  \end{proof}

  \noindent
  Once transitivity is proved, reasoning about \isa{{\isaliteral{5C3C617070726F783E}{\isasymapprox}}} is rather straightforward. For example
  symmetry is a simple consequence.

  \begin{lemma}\label{equsym}
  \isa{{\normalsize{}If\,}\ \ {\isaliteral{5C3C6E61626C613E}{\isasymnabla}}\ {\isaliteral{5C3C7475726E7374696C653E}{\isasymturnstile}}\ t\isaliteral{5C3C5E697375623E}{}\isactrlisub {\isadigit{1}}\ {\isaliteral{5C3C617070726F783E}{\isasymapprox}}\ t\isaliteral{5C3C5E697375623E}{}\isactrlisub {\isadigit{2}}\ {\normalsize \,then\,}\ \ {\isaliteral{5C3C6E61626C613E}{\isasymnabla}}\ {\isaliteral{5C3C7475726E7374696C653E}{\isasymturnstile}}\ t\isaliteral{5C3C5E697375623E}{}\isactrlisub {\isadigit{2}}\ {\isaliteral{5C3C617070726F783E}{\isasymapprox}}\ t\isaliteral{5C3C5E697375623E}{}\isactrlisub {\isadigit{1}}{\isaliteral{2E}{\isachardot}}}
  \end{lemma}

  \begin{proof}
  By induction on \isa{{\isaliteral{5C3C617070726F783E}{\isasymapprox}}}. In the (\isa{{\isaliteral{5C3C617070726F783E}{\isasymapprox}}}-abstraction$_2$) we 
  have \isa{{\isaliteral{5C3C6E61626C613E}{\isasymnabla}}\ {\isaliteral{5C3C7475726E7374696C653E}{\isasymturnstile}}\ {\isaliteral{28}{\isacharparenleft}}a\ b{\isaliteral{29}{\isacharparenright}}\ {\isaliteral{5C3C62756C6C65743E}{\isasymbullet}}\ t\isaliteral{5C3C5E697375623E}{}\isactrlisub {\isadigit{2}}\ {\isaliteral{5C3C617070726F783E}{\isasymapprox}}\ t\isaliteral{5C3C5E697375623E}{}\isactrlisub {\isadigit{1}}} and need to show \isa{{\isaliteral{5C3C6E61626C613E}{\isasymnabla}}\ {\isaliteral{5C3C7475726E7374696C653E}{\isasymturnstile}}\ t\isaliteral{5C3C5E697375623E}{}\isactrlisub {\isadigit{2}}\ {\isaliteral{5C3C617070726F783E}{\isasymapprox}}\ {\isaliteral{28}{\isacharparenleft}}b\ a{\isaliteral{29}{\isacharparenright}}\ {\isaliteral{5C3C62756C6C65743E}{\isasymbullet}}\ t\isaliteral{5C3C5E697375623E}{}\isactrlisub {\isadigit{1}}}. We can 
  do so by inferring \isa{{\isaliteral{5C3C6E61626C613E}{\isasymnabla}}\ {\isaliteral{5C3C7475726E7374696C653E}{\isasymturnstile}}\ {\isaliteral{28}{\isacharparenleft}}b\ a{\isaliteral{29}{\isacharparenright}}{\isaliteral{28}{\isacharparenleft}}a\ b{\isaliteral{29}{\isacharparenright}}\ {\isaliteral{5C3C62756C6C65743E}{\isasymbullet}}\ t\isaliteral{5C3C5E697375623E}{}\isactrlisub {\isadigit{2}}\ {\isaliteral{5C3C617070726F783E}{\isasymapprox}}\ {\isaliteral{28}{\isacharparenleft}}b\ a{\isaliteral{29}{\isacharparenright}}\ {\isaliteral{5C3C62756C6C65743E}{\isasymbullet}}\ t\isaliteral{5C3C5E697375623E}{}\isactrlisub {\isadigit{1}}} using Lemma~\ref{equpi}.
  We can also show \isa{{\isaliteral{5C3C6E61626C613E}{\isasymnabla}}\ {\isaliteral{5C3C7475726E7374696C653E}{\isasymturnstile}}\ {\isaliteral{28}{\isacharparenleft}}b\ a{\isaliteral{29}{\isacharparenright}}{\isaliteral{28}{\isacharparenleft}}a\ b{\isaliteral{29}{\isacharparenright}}\ {\isaliteral{5C3C62756C6C65743E}{\isasymbullet}}\ t\isaliteral{5C3C5E697375623E}{}\isactrlisub {\isadigit{2}}\ {\isaliteral{5C3C617070726F783E}{\isasymapprox}}\ t\isaliteral{5C3C5E697375623E}{}\isactrlisub {\isadigit{2}}} using Lemma~\ref{dsequ}. 
  We can join both facts by transitivity to yield the proof obligation.
  \end{proof}

  \noindent
  To sum up, the neat trick with using \isa{{\isaliteral{5C3C73696D3E}{\isasymsim}}} from \cite{KumarNorrish10} has allowed us
  to give a direct, structural,  proof for equivalence of \isa{{\isaliteral{5C3C617070726F783E}{\isasymapprox}}}. The formalisation of this
  direct proof in Isabelle/HOL is approximately half the size of the formalised proof given 
  in \cite{UrbanPittsGabbay04}.%
\end{isamarkuptext}%
\isamarkuptrue%
\isamarkupsection{An Algorithm for Nominal Unification%
}
\isamarkuptrue%
\begin{isamarkuptext}%
In this section we sketch the algorithm for nominal unification presented in \cite{UrbanPittsGabbay04}. 
  We refer the reader to that paper for full details.

  The purpose of nominal unification algorithm is to calculate substitutions 
  that make terms \isa{{\isaliteral{5C3C617070726F783E}{\isasymapprox}}}-equal.  
  The substitution operation for nominal terms is defined as follows:

  \begin{center}
  \begin{tabular}{r@ {\hspace{2mm}}c@ {\hspace{2mm}}l}
  \isa{{\isaliteral{5C3C7369676D613E}{\isasymsigma}}{\isaliteral{28}{\isacharparenleft}}a{\isaliteral{29}{\isacharparenright}}} & \isa{{\isaliteral{5C3C65717569763E}{\isasymequiv}}} & \isa{a}\\ 
  \isa{{\isaliteral{5C3C7369676D613E}{\isasymsigma}}{\isaliteral{28}{\isacharparenleft}}{\isaliteral{5C3C70693E}{\isasympi}}{\isaliteral{5C3C62756C6C65743E}{\isasymbullet}}X{\isaliteral{29}{\isacharparenright}}} & \isa{{\isaliteral{5C3C65717569763E}{\isasymequiv}}} & 
    $\begin{cases}
       \isa{{\isaliteral{5C3C70693E}{\isasympi}}\ {\isaliteral{5C3C62756C6C65743E}{\isasymbullet}}\ {\isaliteral{5C3C7369676D613E}{\isasymsigma}}{\isaliteral{28}{\isacharparenleft}}X{\isaliteral{29}{\isacharparenright}}}\hspace{5mm}\text{if \isa{X\ {\isaliteral{5C3C696E3E}{\isasymin}}\ dom\ {\isaliteral{5C3C7369676D613E}{\isasymsigma}}}}\\
       \isa{{\isaliteral{5C3C70693E}{\isasympi}}{\isaliteral{5C3C62756C6C65743E}{\isasymbullet}}X}\hspace{5mm}\text{otherwise} 
     \end{cases}$\\ 
  \isa{{\isaliteral{5C3C7369676D613E}{\isasymsigma}}{\isaliteral{28}{\isacharparenleft}}a{\isaliteral{2E}{\isachardot}}t{\isaliteral{29}{\isacharparenright}}} & \isa{{\isaliteral{5C3C65717569763E}{\isasymequiv}}} & \isa{a{\isaliteral{2E}{\isachardot}}{\isaliteral{5C3C7369676D613E}{\isasymsigma}}{\isaliteral{28}{\isacharparenleft}}t{\isaliteral{29}{\isacharparenright}}}\\
  \isa{{\isaliteral{5C3C7369676D613E}{\isasymsigma}}{\isaliteral{28}{\isacharparenleft}}{\isaliteral{5C3C6C616E676C653E}{\isasymlangle}}t\isaliteral{5C3C5E697375623E}{}\isactrlisub {\isadigit{1}}{\isaliteral{2C}{\isacharcomma}}\ t\isaliteral{5C3C5E697375623E}{}\isactrlisub {\isadigit{2}}{\isaliteral{5C3C72616E676C653E}{\isasymrangle}}{\isaliteral{29}{\isacharparenright}}} & \isa{{\isaliteral{5C3C65717569763E}{\isasymequiv}}} & \isa{{\isaliteral{5C3C6C616E676C653E}{\isasymlangle}}{\isaliteral{5C3C7369676D613E}{\isasymsigma}}{\isaliteral{28}{\isacharparenleft}}t\isaliteral{5C3C5E697375623E}{}\isactrlisub {\isadigit{1}}{\isaliteral{29}{\isacharparenright}}{\isaliteral{2C}{\isacharcomma}}\ {\isaliteral{5C3C7369676D613E}{\isasymsigma}}{\isaliteral{28}{\isacharparenleft}}t\isaliteral{5C3C5E697375623E}{}\isactrlisub {\isadigit{2}}{\isaliteral{29}{\isacharparenright}}{\isaliteral{5C3C72616E676C653E}{\isasymrangle}}}\\
  \isa{{\isaliteral{5C3C7369676D613E}{\isasymsigma}}{\isaliteral{28}{\isacharparenleft}}f\ t{\isaliteral{29}{\isacharparenright}}} & \isa{{\isaliteral{5C3C65717569763E}{\isasymequiv}}} & \isa{f\ {\isaliteral{5C3C7369676D613E}{\isasymsigma}}{\isaliteral{28}{\isacharparenleft}}t{\isaliteral{29}{\isacharparenright}}}
  \end{tabular}
  \end{center}

  \noindent
  There are two kinds of problems the nominal unification algorithms solves:

  \begin{center}
  \isa{t\isaliteral{5C3C5E697375623E}{}\isactrlisub {\isadigit{1}}\ {\isaliteral{5C3C617070726F783E}{\isasymapprox}}\isaliteral{5C3C5E7375703E}{}\isactrlsup {\isaliteral{3F}{\isacharquery}}\ t\isaliteral{5C3C5E697375623E}{}\isactrlisub {\isadigit{2}}} \hspace{10mm} \isa{a\ {\isaliteral{23}{\isacharhash}}\isaliteral{5C3C5E7375703E}{}\isactrlsup {\isaliteral{3F}{\isacharquery}}\ t}
  \end{center}

  \noindent
  The first are called equational problems, the second freshness problems.
  Their respective interpretation is ``can the terms \isa{t\isaliteral{5C3C5E697375623E}{}\isactrlisub {\isadigit{1}}} and \isa{t\isaliteral{5C3C5E697375623E}{}\isactrlisub {\isadigit{2}}} be made
  equal according to \isa{{\isaliteral{5C3C617070726F783E}{\isasymapprox}}}?'' and ``can the atom \isa{a} be made fresh for
  \isa{t} according to \isa{{\isaliteral{23}{\isacharhash}}}?''. A solution for each kind of problems is a pair \isa{{\isaliteral{28}{\isacharparenleft}}{\isaliteral{5C3C6E61626C613E}{\isasymnabla}}{\isaliteral{2C}{\isacharcomma}}\ {\isaliteral{5C3C7369676D613E}{\isasymsigma}}{\isaliteral{29}{\isacharparenright}}} 
  consisting of a freshness environment and a substitution such that

  \begin{center}
  \isa{{\isaliteral{5C3C6E61626C613E}{\isasymnabla}}\ {\isaliteral{5C3C7475726E7374696C653E}{\isasymturnstile}}\ {\isaliteral{5C3C7369676D613E}{\isasymsigma}}{\isaliteral{28}{\isacharparenleft}}t\isaliteral{5C3C5E697375623E}{}\isactrlisub {\isadigit{1}}{\isaliteral{29}{\isacharparenright}}\ {\isaliteral{5C3C617070726F783E}{\isasymapprox}}\ {\isaliteral{5C3C7369676D613E}{\isasymsigma}}{\isaliteral{28}{\isacharparenleft}}t\isaliteral{5C3C5E697375623E}{}\isactrlisub {\isadigit{2}}{\isaliteral{29}{\isacharparenright}}} \hspace{10mm} \isa{{\isaliteral{5C3C6E61626C613E}{\isasymnabla}}\ {\isaliteral{5C3C7475726E7374696C653E}{\isasymturnstile}}\ a\ {\isaliteral{23}{\isacharhash}}\ {\isaliteral{5C3C7369676D613E}{\isasymsigma}}{\isaliteral{28}{\isacharparenleft}}t{\isaliteral{29}{\isacharparenright}}}
  \end{center}

  \noindent
  hold. Note the difference with first-order unification and higher-order pattern unification
  where a solution consists of a substitution only. An example where nominal unification
  calculates a non-trivial
  freshness environment is the equational problem

  \begin{center}
  \isa{a{\isaliteral{2E}{\isachardot}}X\ {\isaliteral{5C3C617070726F783E}{\isasymapprox}}\isaliteral{5C3C5E7375703E}{}\isactrlsup {\isaliteral{3F}{\isacharquery}}\ b{\isaliteral{2E}{\isachardot}}X}
  \end{center}

  \noindent
  which is solved by the solution \isa{{\isaliteral{28}{\isacharparenleft}}{\isaliteral{7B}{\isacharbraceleft}}a\ {\isaliteral{23}{\isacharhash}}\ X{\isaliteral{2C}{\isacharcomma}}\ b\ {\isaliteral{23}{\isacharhash}}\ X{\isaliteral{7D}{\isacharbraceright}}{\isaliteral{2C}{\isacharcomma}}\ {\isaliteral{5B}{\isacharbrackleft}}{\isaliteral{5D}{\isacharbrackright}}{\isaliteral{29}{\isacharparenright}}}.
  Solutions in nominal unification can be ordered so that the unification 
  algorithm produces always most general solutions.  This ordering is defined
  very similar to the standard ordering in first-order unification.

  The nominal unification algorithm
  in \cite{UrbanPittsGabbay04} is defined in the usual style of rewriting rules that transform
  sets of unification problems to simpler ones calculating a substitution and
  freshness environment on the way. The transformation rule for pairs is

  \begin{center}
  \isa{{\isaliteral{7B}{\isacharbraceleft}}{\isaliteral{5C3C6C616E676C653E}{\isasymlangle}}t\isaliteral{5C3C5E697375623E}{}\isactrlisub {\isadigit{1}}{\isaliteral{2C}{\isacharcomma}}\ t\isaliteral{5C3C5E697375623E}{}\isactrlisub {\isadigit{2}}{\isaliteral{5C3C72616E676C653E}{\isasymrangle}}\ {\isaliteral{5C3C617070726F783E}{\isasymapprox}}\isaliteral{5C3C5E7375703E}{}\isactrlsup {\isaliteral{3F}{\isacharquery}}\ {\isaliteral{5C3C6C616E676C653E}{\isasymlangle}}s\isaliteral{5C3C5E697375623E}{}\isactrlisub {\isadigit{1}}{\isaliteral{2C}{\isacharcomma}}\ s\isaliteral{5C3C5E697375623E}{}\isactrlisub {\isadigit{2}}{\isaliteral{5C3C72616E676C653E}{\isasymrangle}}{\isaliteral{2C}{\isacharcomma}}\ {\isaliteral{5C3C646F74733E}{\isasymdots}}{\isaliteral{7D}{\isacharbraceright}}\ {\isaliteral{5C3C4C6F6E6772696768746172726F773E}{\isasymLongrightarrow}}\ {\isaliteral{7B}{\isacharbraceleft}}t\isaliteral{5C3C5E697375623E}{}\isactrlisub {\isadigit{1}}\ {\isaliteral{5C3C617070726F783E}{\isasymapprox}}\isaliteral{5C3C5E7375703E}{}\isactrlsup {\isaliteral{3F}{\isacharquery}}\ s\isaliteral{5C3C5E697375623E}{}\isactrlisub {\isadigit{1}}{\isaliteral{2C}{\isacharcomma}}\ t\isaliteral{5C3C5E697375623E}{}\isactrlisub {\isadigit{2}}\ {\isaliteral{5C3C617070726F783E}{\isasymapprox}}\isaliteral{5C3C5E7375703E}{}\isactrlsup {\isaliteral{3F}{\isacharquery}}\ s\isaliteral{5C3C5E697375623E}{}\isactrlisub {\isadigit{2}}{\isaliteral{2C}{\isacharcomma}}\ {\isaliteral{5C3C646F74733E}{\isasymdots}}{\isaliteral{7D}{\isacharbraceright}}}
  \end{center}

  \noindent
  There are two rules for abstractions depending on whether or not the binders agree.

  \begin{center}
  \begin{tabular}{l}
  \isa{{\isaliteral{7B}{\isacharbraceleft}}a{\isaliteral{2E}{\isachardot}}t\ {\isaliteral{5C3C617070726F783E}{\isasymapprox}}\isaliteral{5C3C5E7375703E}{}\isactrlsup {\isaliteral{3F}{\isacharquery}}\ a{\isaliteral{2E}{\isachardot}}s{\isaliteral{2C}{\isacharcomma}}\ {\isaliteral{5C3C646F74733E}{\isasymdots}}{\isaliteral{7D}{\isacharbraceright}}\ {\isaliteral{5C3C4C6F6E6772696768746172726F773E}{\isasymLongrightarrow}}\ {\isaliteral{7B}{\isacharbraceleft}}t\ {\isaliteral{5C3C617070726F783E}{\isasymapprox}}\isaliteral{5C3C5E7375703E}{}\isactrlsup {\isaliteral{3F}{\isacharquery}}\ s{\isaliteral{2C}{\isacharcomma}}\ {\isaliteral{5C3C646F74733E}{\isasymdots}}{\isaliteral{7D}{\isacharbraceright}}}\\
  \isa{{\isaliteral{7B}{\isacharbraceleft}}a{\isaliteral{2E}{\isachardot}}t\ {\isaliteral{5C3C617070726F783E}{\isasymapprox}}\isaliteral{5C3C5E7375703E}{}\isactrlsup {\isaliteral{3F}{\isacharquery}}\ b{\isaliteral{2E}{\isachardot}}s{\isaliteral{2C}{\isacharcomma}}\ {\isaliteral{5C3C646F74733E}{\isasymdots}}{\isaliteral{7D}{\isacharbraceright}}\ {\isaliteral{5C3C4C6F6E6772696768746172726F773E}{\isasymLongrightarrow}}\ {\isaliteral{7B}{\isacharbraceleft}}t\ {\isaliteral{5C3C617070726F783E}{\isasymapprox}}\isaliteral{5C3C5E7375703E}{}\isactrlsup {\isaliteral{3F}{\isacharquery}}\ {\isaliteral{28}{\isacharparenleft}}a\ b{\isaliteral{29}{\isacharparenright}}\ {\isaliteral{5C3C62756C6C65743E}{\isasymbullet}}\ s{\isaliteral{2C}{\isacharcomma}}\ a\ {\isaliteral{23}{\isacharhash}}\isaliteral{5C3C5E7375703E}{}\isactrlsup {\isaliteral{3F}{\isacharquery}}\ s{\isaliteral{2C}{\isacharcomma}}\ {\isaliteral{5C3C646F74733E}{\isasymdots}}{\isaliteral{7D}{\isacharbraceright}}}
  \end{tabular}
  \end{center}

  \noindent
  One rule that is also interesting is for unifying two suspensions with
  the same variable

  \begin{center}
  \isa{{\isaliteral{7B}{\isacharbraceleft}}{\isaliteral{5C3C70693E}{\isasympi}}{\isaliteral{5C3C62756C6C65743E}{\isasymbullet}}X\ {\isaliteral{5C3C617070726F783E}{\isasymapprox}}\isaliteral{5C3C5E7375703E}{}\isactrlsup {\isaliteral{3F}{\isacharquery}}\ {\isaliteral{5C3C70693E}{\isasympi}}{\isaliteral{27}{\isacharprime}}{\isaliteral{5C3C62756C6C65743E}{\isasymbullet}}X{\isaliteral{2C}{\isacharcomma}}{\isaliteral{5C3C646F74733E}{\isasymdots}}{\isaliteral{7D}{\isacharbraceright}}\ {\isaliteral{5C3C4C6F6E6772696768746172726F773E}{\isasymLongrightarrow}}\ {\isaliteral{7B}{\isacharbraceleft}}a\ {\isaliteral{23}{\isacharhash}}\isaliteral{5C3C5E7375703E}{}\isactrlsup {\isaliteral{3F}{\isacharquery}}\ X\ {\isaliteral{7C}{\isacharbar}}\ a\ {\isaliteral{5C3C696E3E}{\isasymin}}\ ds\ {\isaliteral{5C3C70693E}{\isasympi}}\ {\isaliteral{5C3C70693E}{\isasympi}}{\isaliteral{27}{\isacharprime}}{\isaliteral{7D}{\isacharbraceright}}\ {\isaliteral{5C3C756E696F6E3E}{\isasymunion}}\ {\isaliteral{7B}{\isacharbraceleft}}{\isaliteral{5C3C646F74733E}{\isasymdots}}{\isaliteral{7D}{\isacharbraceright}}}
  \end{center}

  \noindent
  What is interesting about nominal unification is that it never needs to
  create fresh names. As can be seen from the abstraction rules, no new name needs
  to be introduced in order to unify abstractions. It is the case that all atoms
  in a solution, occur already in the original problem. This has the attractive 
  consequence that nominal unification can dispense with any new-name-generation
  facility. This makes it easy to implement and reason about the nominal
  unification algorithm. Clearly, however, the running time of the algorithm using the
  rules sketched above is exponential in the worst-case, just like the
  simple-minded first-order unification algorithm without sharing.%
\end{isamarkuptext}%
\isamarkuptrue%
\isamarkupsection{Applications and Complexity of Nominal Unification%
}
\isamarkuptrue%
\begin{isamarkuptext}%
Having designed a new algorithm for unification, it is an obvious step to include it into 
  a logic programming language. This has been studied in the work about 
  $\alpha$Prolog \cite{CheneyUrban04} and
  $\alpha$Kanren \cite{ByrdFriedman07}. The latter is a system implemented on top of
  Scheme and is more sophisticated than the former.
  The point of these variants of Prolog is that
  they allow one to implement inference rule systems in a very concise and declarative
  manner. For example the typing rules for simply-typed lambda-terms shown in the
  Introduction can be implemented
  in $\alpha$Prolog as follows:

  \begin{center}
  \begin{tabular}{@ {}l}
  {\hspace{4.3mm}\tt type (Gamma, var(X), T) :- member (X,T) Gamma.}\\[2mm]
  {\hspace{4.3mm}\tt type (Gamma, app(M,N), T2) :-}\\[0.0mm]
  {\hspace{1.2cm}\tt       type (Gamma, M, arrow(T1, T2)), type (Gamma, N, T1).}\\[2mm]
  {\hspace{4.3mm}\tt type (Gamma, \colorbox{mygrey}{lam(x.M)}, arrow(T1, T2)) /
  \colorbox{mygrey}{x \# Gamma} :-}\\
  {\hspace{1.2cm}\tt type ((x,T1)::Gamma, M, T2).}\\[4mm]
  {\hspace{4.3mm}\tt member X X::Tail.}\\
  {\hspace{4.3mm}\tt member X Y::Tail :- member X Tail.}\\
  \end{tabular}
  \end{center}

  \noindent
  The shaded boxes show two novel features of $\alpha$Prolog.  Abstractions can
  be written as {\tt x.$(-)$}; but note that the binder {\tt x} can also occur
  as a ``non-binder'' in the body of clauses---just as in the clauses on
  ``paper.''  The side-condition {\tt x\,\#\,Gamma} ensures that {\tt x} is not
  free in any term substituted for {\tt Gamma}. The novel features of
  $\alpha$Prolog and $\alpha$Kanren can be appreciated when considering that similarly simple
  implementations in ``vanilla'' Prolog (which, surprisingly, one can find in textbooks 
  \cite{Mitchell03}) are
  \emph{incorrect}, as they give types to untypable lambda-terms. An 
  simple implementation of a first-order theorem prover in $\alpha$Kanren
  has been given in \cite{alphaleantap08}.

  When implementing a logic programming language based on nominal unification
  it becomes important to answer the question about its complexity.  Surprisingly,
  this turned out to be a difficult question. Surprising because nominal unification,
  like first-order unification, uses simple rewrite rules defined over first-order 
  terms and uses a substitution operation that is a simple replacement of terms 
  for variables. One would hope the techniques from efficient first-order unification
  algorithms carry over to nominal unification. This is unfortunately only partially
  the case. Quadratic algorithms for nominal unification were obtained by 
  Calves and Fernandez \cite{CalvesFernandez08,Calves10} and independently 
  by Levy and Villaret \cite{levyvillaret10}. These are the best bounds we have 
  for nominal unification so far.%
\end{isamarkuptext}%
\isamarkuptrue%
\isamarkupsection{Conclusion%
}
\isamarkuptrue%
\begin{isamarkuptext}%
Nominal unification was introduced in \cite{UrbanPittsGabbay04}. It unifies terms
  involving binders modulo a notion of alpha-equivalence. In this way it is more
  powerful than first-order unification, but is conceptually much simpler than
  higher-order pattern unification. Unification algorithms are often critical
  infrastructure in theorem provers. Therefore it is important to formalise
  these algorithms in order to ensure correctness. Nominal unification has
  been formalised twice, once in \cite{UrbanPittsGabbay04} in Isabelle/HOL 
  and another in \cite{KumarNorrish10} in HOL4. The latter formalises a more 
  efficient version of nominal unification based on triangular substitutions. 
  The main purpose of this paper is to simplify the transitivity proof for \isa{{\isaliteral{5C3C617070726F783E}{\isasymapprox}}}. 
  This in turn simplified the formalisation in Isabelle/HOL.

  There have been several fruitful avenues of research that use nominal unification
  as basic building block. For example the work  on $\alpha$LeanTap \cite{alphaleantap08}. 
  There have also been several works that go beyond the limitation
  of nominal unification where bound names are restricted to be constant symbols that are not 
  substitutable \cite{PittsAM:residb,PittsAM:nomel}.%
\end{isamarkuptext}%
\isamarkuptrue%
\isadelimtheory
\endisadelimtheory
\isatagtheory
\endisatagtheory
{\isafoldtheory}%
\isadelimtheory
\endisadelimtheory
\end{isabellebody}%


\bibliographystyle{plain}
\bibliography{root}

\end{document}